\documentclass[12pt]{article}
\usepackage{amssymb,amsmath,epsfig,graphicx}
\begin{document}
\title{\bf Physical Attributes of Anisotropic Compact Stars in $f(R,G)$ Gravity}
\author{M. Farasat Shamir
\thanks{farasat.shamir@nu.edu.pk} and Saeeda Zia \thanks{saeeda.zia@nu.edu.pk} \\\\
Department of Sciences and Humanities, \\National University of
Computer and Emerging Sciences,\\ Lahore Campus, Pakistan.}

\date{}
\maketitle

\date{}

\maketitle
\begin{abstract}
Modified gravity is one of the potential candidates to explain the accelerated expansion of the universe. Current study highlights the materialization of anisotropic compact stars in the context of $f(R,G)$ theory of gravity. In particular, to gain insight in the physical behavior of three stars namely, Her $X1$, SAX J $1808$-$3658$ and 4U $1820$-$30$, energy density, and radial and tangential pressures are calculated. The $f(R,G)$ gravity model is split into a Starobinsky like $f(R)$ model and a power law $f(G)$ model. The main feature of the work is a $3$-dimensional graphical analysis in which, anisotropic measurements, energy conditions and stability attributes of these stars are discussed. It is shown that all three stars behave as usual for positive values of $f(G)$ model parameter $n$.
\end{abstract}


\section{Introduction}
The spatial behavior of constituents of the universe is a complex phenomena. Visual observation of the remote universe enlightens us about its continuous expansion, and accentuates to find alternative rationale which may be helpful to explain the phenomena of  expansion of the universe \cite{1}. So, these alternative models are known currently as modified theories of general relativity (GR), some of which are $f(R)$, $f(G)$,  $f(R,T)$, $f(G,T)$ and $f(R,G)$ theories of gravity, where $R$, $G$, and $T$ denotes Ricci scalar, the Gauss-Bonnet invariant and the trace of the energy momentum tensor respectively. The theory of GR explains the cosmological phenomena in weak field regimes while some modifications are required to address the strong fields in the scenario of continuous expansion of universe. The same thought was provoked and one of the modified theories, $f(R)$ gravity,  was proposed in $1970$ by Buchdahl \cite{2}. Hydrostatic equilibrium and stellar structure in $f(R)$ gravity have been investigated by considering the Lan´e-Emden equation \cite{2a}. Some finite time future singularities in modified gravity were described with their solution using the addition of higher derivative gravitational invariants \cite{2b}.  Harko  \cite{5} presented modified $f(R,T)$ theory in $2011$ by involving both matter and curvature terms. Furthermore, the next modification to the GR was Gauss-Bonnet gravity \cite{02}, also known as Einstein-Gauss-Bonnet gravity, which includes the Gauss-Bonnet term.
\begin{equation}
G\equiv R^2-4R_{\mu\nu}R^{\mu\nu}+R_{\mu\nu\theta\phi}R^{\mu\nu\theta\phi},
\end{equation}
where $R_{\mu\nu}$ and $R_{\mu\nu\theta\phi}$ are Ricci tensor and the Riemann tensor respectively.
$f(G)$ gravity \cite{3} is the further generalization of Gauss-Bonnet gravity. It has been shown that this kind of generalization may naturally approach an operative cosmological constant, quintessence and phantom cosmic acceleration \cite{3a}. Recently, another alternative theory with the title of $f(G,T)$ gravity has been proposed by Sharif and Ikram \cite{6}. They studied the energy conditions for the well-known Friedmann -Robertson- Walker space time. So far some interesting work has been done in this modified theory \cite{shamir.ahmad}-\cite{sharif.ayesha2}.
Similarly, $f(\mathcal{T})$ gravity is also a modified theory, which has attracted the attention of theoretical physicists gain insight of acceleration and regular thermal expansion of late-time universe by assuming an alternative of cosmological constant, where $f(\mathcal{T})$ is a function of the torsion scalar \cite{4}. An endeavor was made by combining $R$ and $G$ in a bivariate function $f(R,G)$ \cite{7}-\cite{14}. The function $f(R,G)$ provides a foundation for the double inflationary scenario \cite{15}. It has been shown that $f(R,G)$  theory is consistent with the observational data \cite{16,17}. Besides its stability, this theory is perfectly suitable to explain the accelerating waves of the celestial bodies as well as the phantom divide line crossing and transition from acceleration to deceleration phases. An important feature of $f(R,G)$ gravity is that it reduces the risk of ghost contributions and gravitational action is regularized because of the term $G$ \cite{18,19}. Thus, modified theories of gravity seem interesting in explaining the universe in different cosmological contexts \cite{19a}.\\
Study of compact stars has always been a strong topic of research \cite{20}-\cite{32c}. Different properties like mass, radius and moment of inertia  of neutron stars are studied and comparison has been developed with GR and alternative theories of gravity \cite{32d}. Some investigations of the structure of slowly rotating neutron stars in $R^2$ gravity have been done using two different hadronic parameters and a strange matter equation of state (EOS) parameter \cite{32e}.
The mass-ratio for compact objects in the presence of cosmological constant has been derived \cite{24}.
In this article, we are interested in studying the compact stars and in examining their physical behavior using standard models of spherically symmetric space time by finding exact solution of field equations. We study the stability of anisotropic compact stars using generalized modified Gauss-Bonnet $f(R,G)$ gravity.
The arrangement of this paper is as follows: In the second section, we find expressions for energy density, radial pressure and tangential pressure for anisotropic matter source using spherically symmetric metric in $f(R,G)$ gravity. Section \textbf{3} comprises the analysis of behavior of EOS parameters, anisotropy measure, the matching conditions, the energy bounds, Tolman Oppenheimer Volkoff (TOV) equation and stability analysis of three compact stars namely, Her $X1$, SAX J $1808$-$3658$ and 4U $1820$-$30$. The last section ends with the conclusion and summery of the paper.

\section{Anisotropic Matter Configuration in $f(R,G)$ Gravity}

We start with the action of modified Gauss-Bonnet gravity \cite{33}
\begin{eqnarray}
S=\frac{1}{2K}\int d^4x\sqrt{-g}f(R,G)+S_M(g^{\mu\nu},\psi).\label{1}
\end{eqnarray}
Varying action (\ref{1}) with respect to metric tensor yields the following modified field equations \cite{17}
\begin{eqnarray}\label{2}
R_{\mu\nu}-\frac{1}{2}g_{\mu\nu}R&=&\kappa T^{(matt)}_{\mu\nu}+\nabla_\mu\nabla_\nu f_R-g_{\mu\nu}\Box f_R+2R\nabla_\mu\nabla_\nu f_G-2g_{\mu\nu}R\Box f_G\nonumber
\\&& -4R^\alpha_\mu\nabla_\alpha\nabla_\nu f_G
-4R^\alpha_\nu\nabla_\alpha\nabla_\mu f_G+4R_{\mu\nu}\Box f_G+4g_{\mu\nu}R^{\theta\phi}\nabla_\theta\nabla_\phi f_G\nonumber \\&&+4R_{\mu\theta\phi\nu}\nabla^\theta\nabla^\phi f_G-\frac{1}{2}g_{\mu\nu}V+(1-f_R)G_{\mu\nu},
\end{eqnarray}
where $f_R$ and $f_G$ are partial derivatives with respect to $R$ and $G$ respectively,
\begin{eqnarray}
V\equiv f_RR+f_GG-f(R,G),
\end{eqnarray}
and $T_{\mu\nu}^{(matt)}$ describes the ordinary matter.
The most general spherically symmetric  space time is \cite{34}
\begin{eqnarray}\label{3}
ds^{2}=e^{a(r)}dt^{2}-e^{b(r)}dr^{2}-r^{2}{(d\theta^{2}+\sin^{2}\theta d\phi^{2})}.
\end{eqnarray}
The energy-momentum tensor in case of anisotropic fluid is given by
\begin{equation}\label{4}
T_{\alpha\beta}^{m}=(\rho+p_t)u_\alpha u_\beta-p_tg_{\alpha\beta}+(p_r-p_t)v_\alpha v_\beta,
\end{equation}
where $u_\alpha=e^{a/2} \delta_\alpha^0$, $v_\alpha=e^{b/2}\delta_\alpha^1$ are four velocities. Radial pressure and tangential pressures are $p_r$ and $p_t$ respectively, and the energy density is denoted by $\rho$. Using equations (\ref{3}) and (\ref{4}) in field equations (\ref{2}), and after some manipulations we obtain,
\begin{eqnarray}\label{5}
\rho&=&-e^{-b} f''_{1R}-\frac{4e^{-2b}}{r^2}(a'r-b'r-e^b+4)f''_{2G}-e^{-b}(\frac{b'}{2}+
\frac{2}{r})f'_{1R}+e^{-2b}(a''a'+\nonumber\\
&& a''b'-a''-a'^2+a'b'+\frac{a'^3}{2}-\frac{4a'}{r}-\frac{4b'e^b}{r^2}+\frac{8a''}{r}+\frac{6a'^2}{r}
-\frac{13b'}{r^2}-\frac{2b'^2}{r}-\nonumber\\
&&\frac{17a'}{r^2}-\frac{18e^b}{r^3}+\frac{18}{r^3})f'_{2G}+
\frac{e^{-b}}{r^2}(\frac{a''r^2}{2}+\frac{a'^2r^2}{2}-\frac{a'b'r^2}{4}+a'r)f_{1R}+\frac{e^{-2b}}{r^2}\nonumber\\
&&\{(1-e^b)(a'^2+2a''-a'b')-2a'b')\}f_{2G}-\frac{f}{2},
\end{eqnarray}
\begin{eqnarray}\label{6}
p_r&=&e^{-b}\big(\frac{a'}{2}+b'+\frac{2}{r}\big)f'_{1R}+e^{-2b}\{(2a''b'+a'^2b'-a'b'^2+
\frac{4a'b'}{r}+\frac{8a'}{r^2}-\frac{4b'^2}{r}-\nonumber\\
&&\frac{4a'e^b}{r^2}-\frac{4b'e^b}{r^2} -\frac{8e^b}{r^3}+\frac{2a'}{r^2}+\frac{4b'}{r^2}+\frac{8}{r^3})-r(2a''b'+a'^2b'-a'b'^2-
\frac{4b'^2}{r})\}f'_{2G}\nonumber\\&&-\frac{e^{-b}}{r^2}(\frac{a''r^2}{2}+\frac{a'^2r^2}{4}-
\frac{a'b'r^2}{4}-b'r)f_{1R}-\frac{e^{-2b}}{r^2}\{(1-e^b)(a'^2+2a''-a'b')-\nonumber\\&& 2a'b'\}f_{2G}+\frac{f}{2},
\end{eqnarray}
\begin{eqnarray}\label{7}
p_t&=&e^{-b} f''_{1R}+e^{-2b}(2a''+a'^2-a'b'+\frac{2a'}{r}-\frac{2b'}{r})f''_{2G}+e^{-b}(\frac{a'}{2}+
\frac{b'}{2}+\frac{1}{r})f'_{1R}\nonumber
\\&&-\frac{1}{2r^3}\{e^{-b}(2b'^2r^2+4b'r-4a'^2r^2-16a'r-24-2a'a''r^3+a'^2b'r^3-a'^3r^3)\nonumber
\\&&+32-8e^b\}f'_{2G}-\frac{e^{-b}}{r^2}(\frac{a'r}{2}-\frac{b'r}{2}-e^b+1)f_{1R}-\frac{e^{-2b}}{r^2}\{(1-e^b)(a'^2+2a''-a'b')\nonumber
\\&&-2a'b'\}f_{2G}+\frac{f}{2}.
\end{eqnarray}
The set of three equations (\ref{5})-(\ref{7}) involves five unknown functions $\rho, p_r, p_t, a, b$. Moreover, the equations are too much complicated and highly non-linear due to the involved bivariate function $f(R,G)$. So following Krori and Barua \cite{34}, we choose $a(r)=Br^{2}+C$, $ b(r)=Ar^{2}$, $A$, $B$, and $C$ are constants. These constants will be determined by using some physical assumptions. For the present analysis, we propose $f(R,G)=f_1(R)+f_2(G)$. We further consider the Starobinsky like model $f_1(R)=R+\lambda R^2$, where $\lambda$ is an arbitrary constant and we take $f_2(G)=G^n$, $n\neq 0$. Also $f_{1R}=\frac{df_1}{dR}$, $f_{2G}=\frac{df_2}{dG}$ and a prime denotes the derivatives with respect to the radial coordinate.
Using these assumptions, Eqs. (\ref{5})-(\ref{7}) take the form
\begin{eqnarray}
\rho&=&\frac{1}{2r^4}\bigg[-8^nr^4\bigg(-\frac{Be^{-2Ar^2}(-1+3Ar^2-Br^2+e^{Ar^2}(1-Ar^2+Br^2))}{r^2}\bigg)^n+\nonumber
\\&&8^nr^4\bigg(-\frac{Be^{-2Ar^2}(-1+3Ar^2-Br^2+e^{Ar^2}(1-Ar^2+Br^2))}{r^2}\bigg)^n+\nonumber
\\&&\frac{8^n(-1+n)nr^2}{(-1+3Ar^2-Br^2+e^{Ar^2}(1-Ar^2+Br^2))^2}\times\nonumber
\\&&\bigg(-\frac{Be^{-2Ar^2}(-1+3Ar^2-Br^2+e^{Ar^2}(1-Ar^2+Br^2))}{r^2}\bigg)^n\nonumber
\\&&(-1+6A^2r^4-2A(r^2+Br^4)+e^{Ar^2}(1-A^2r^4+A(r^2+Br^4)))(-e^{Ar^2}(9+4Ar^2)+\nonumber
\\&&r^2(-4A^2r^2+B(-9-5r+14Br^2-2Br^3+2B^2r^4)+A(-13+2Br^2(1+r))))+\nonumber
\\&&\frac{2^{1+3n}(-1+n)nr^2(-4+e^{Ar^2}+2Ar^2-2Br^2)}{B(-1+3Ar^2-Br^2+e^{Ar^2}(1-Ar^2+Br^2))^3}\times\nonumber
\\&&\bigg(-\frac{Be^{-2Ar^2}(-1+3Ar^2-Br^2+e^{Ar^2}(1-Ar^2+Br^2))}{r^2}\bigg)^n\times\nonumber
\\&&(-1-19Ar^2+3Br^2+28A^2r^4-12ABr^4+30A^3r^6-4A^2Br^6-72A^4r^8+48A^3Br^8-\nonumber
\end{eqnarray}
\begin{eqnarray}
&&8A^2B^2r^8+2n(1-6A^2r^4+2A(r^2+Br^4))^2+e^{2Ar^2}(-1+3Br^2-2A^4r^8+A^3r^6\times\nonumber
\\&&(3+4Br^2)-2A^2r^4(-2+Br^2+B^2r^4)-Ar^2(8+6Br^2+B^2r^4)+2n(1-A^2r^4+\nonumber
\\&&A(r^2+Br^4))^2+e^{Ar^2}(2-6Br^2+18A^4r^8+3Ar^2(3+Br^2)^2-3A^3r^6(5+8Br^2)+\nonumber
\\&&6A^2r^4(-7+B^2r^4)-4n(1+6A^4r^8+3A(r^2+Br^4)+A^2r^4(-5+4Br^2+2B^2r^4)-\nonumber
\\&&8A^3(r^6+Br^8))))+16e^{-2Ar^2}(-3+3e^{Ar^2}-B^2r^4+2A^3r^6(2+Br^2)-\nonumber
\\&&A^2r^4(4+11Br^2+2B^2r^4)+Ar^2(-3+4Br^2+5B^2r^4))\lambda-\nonumber
\\&&16e^{-2Ar^2}(2+Ar^2)(-1+e^{Ar^2}+B^2r^4(2+Br^2)-A(r^2+4Br^4+B^2r^6))\lambda+\nonumber
\\&&2e^{-2Ar^2}(-1+e^{Ar^2}-3Br^2-B^2r^4+Ar^2(2+Br^2))\times\nonumber
\\&&(e^{Ar^2}(r^2-2\lambda)+2(1+3Br^2+B^2r^4-Ar^2(2+Br^2))\lambda)+2Be^{-2Ar^2}r^2\nonumber
\\&&(3-Ar^2+2Br^2)\big(e^{Ar^2}(r^2-4\lambda)+4(1+3Br^2+B62r^4-Ar^2(2+Br^2))\lambda\big)\bigg],\nonumber\\\label{8}
\end{eqnarray}
\begin{eqnarray}
p_r&=&\frac{1}{2}\bigg[8^n\bigg(-\frac{Be^{-2Ar^2}(-1+3Ar^2-Br^2+e^{Ar^2}(1-Ar^2+Br^2))}{r^2}\bigg)^n-\nonumber
\\&&8^nn\bigg(-\frac{Be^{-2Ar^2}(-1+3Ar^2-Br^2+e^{Ar^2}(1-Ar^2+Br^2))}{r^2}\bigg)^n-\nonumber
\\&&(2^{1+3n}(-1+n)n\bigg(-\frac{Be^{-2Ar^2}(-1+3Ar^2-Br^2+e^{Ar^2}(1-Ar^2+Br^2))}{r^2}\bigg)^n\times\nonumber
\\&&(-2-5Br^2-2A^2(-1+r)r^4(2+Br^2)+2e^{Ar^2}(1+Ar^2+Br^2)+2Ar^2\nonumber
\\&&(-1+B(-3+r)r^2+B^2(-1+r)r^4))(-1+6A^2r^4-2A(r^2+Br^4)+e^{Ar^2}\nonumber
\\&&(1-A^2r^4+A(r^2+Br^4))))\times\frac{1}{(Br^2(1-3Ar^2+Br^2+e^{Ar^2}(-1+Ar^2-Br^2))^2}\nonumber
\\&&+\frac{16e^{-2Ar^2}}{r^4}\times(2+2Ar^2+Br^2)(-1+e^{Ar^2}+B^2r^4+A^2r^4(2+Br^2)-\nonumber
\\&&A(r^2+4Br^4+B^2r^6))\lambda-\frac{2}{r^4}e^{-2Ar^2}(-1+e^{Ar^2}-3Br^2-B^2r^4+Ar^2(2+Br^2))\times\nonumber
\\&&(e^{Ar^2}(r^2-2\lambda)+2(1+3Br^2+B^2r^4-Ar^2(2+Br^2))\lambda)+2e^{-2Ar^2}\times\nonumber
\\&&\frac{(-B(1+Br^2)+A(2+Br^2))(e^{Ar^2}(r^2-4\lambda)+4(1+3Br^2+B^2r^4-Ar^2(2+Br^2))\lambda}{r^2}\bigg],\nonumber\\\label{9}
\end{eqnarray}
\begin{eqnarray}
p_t&=&\frac{1}{r^6}\bigg[2^{-1+3n}r^6\bigg(-\frac{Be^{-2Ar^2}(-1+3Ar^2-Br^2+e^{Ar^2}(1-Ar^2+Br^2))}{r^2}\bigg)^n-\nonumber
\\&&2^{-1+3n}nr^6\bigg(-\frac{Be^{-2Ar^2}(-1+3Ar^2-Br^2+e^{Ar^2}(1-Ar^2+Br^2))}{r^2}\bigg)^n+\nonumber
\\&&\frac{8^ne^{Ar^2}(-1+n)nr^4}{B(1-3Ar^2+Br^2+e^{Ar^2}(-1+Ar^2-Br^2))^2}\times\nonumber
\\&&\bigg(-\frac{Be^{-2Ar^2}(-1+3Ar^2-Br^2+e^{Ar^2}(1-Ar^2+Br^2))}{r^2}\bigg)^n-\nonumber
\\&&(3-4e^{Ar^2}+e^{2Ar^2}+4Br^2-A^2r^4+3B^2r^4+B^3r^6-A(r^2+B^2r^6))\times\nonumber
\\&&(-1+6A^2r^4-2A(r^2+Br^4)+e^{Ar^2}(1-A^2r^4+A(r^2+Br^4)))+\nonumber
\\&&\frac{8^n(-1+n)nr^6(A+ABr^2-B(2+Br^2))}{B(-1+3Ar^2-Br^2+e^{Ar^2}(1-Ar^2+Br^2))^3}\times\nonumber
\\&&\bigg(-\frac{Be^{-2Ar^2}(-1+3Ar^2-Br^2+e^{Ar^2}(1-Ar^2+Br^2))}{r^2}\bigg)^n\nonumber
\\&&(-1-19Ar^2+3Br^2+28A^2r^4-12ABr^4+30A^3r^6-4A^2Br^6-2AB^2r^6-72A^4r^8+\nonumber
\\&&48A^3Br^8-8A^2B^2r^8+2n(1-6A^2r^4+2A(r^2+Br^4))^2+e^{2Ar^2}(-1+3Br^2-2A^4r^8\nonumber
\\&&+A^3r^6(3+4Br^2)-2A^2r^4(-2+Br^2+B^2r^4)-Ar^2(8+6Br^2+B^2r^4)+\nonumber
\\&&2n(1-A^2r^4+A(r^2+Br^4))^2)+e^{Ar^2}(2-6Br^2+18A^4r^8+3Ar^2(3+Br^2)^2-\nonumber
\\&&3A^3r^6(5+8Br^2)+6A^2r^4(-7+B^2r^4)-4n(1+6A^4r^8+3A(r^2+Br^4)+\nonumber
\\&&A^2r^4(-5+4Br^2+2B^2r^4)-8A^3(r^6+Br^8))))-8e^{-2Ar^2}r^2(-3+3e^{Ar^2}-B^2r^4+\nonumber
\\&&2A^3r^6(2+Br^2)-A^2r^4(4+11Br^2+2B^2r^4)+Ar^2(-3+4Br^2+5B^2r^4))\lambda+8e^{-2Ar^2}r^2\nonumber
\\&&(1+Ar^2+Br^2)(-1+e^{Ar^2}+B^2r^4+A^2r^4(2+Br^2)-A(r^2+4Br^4+B^2r^6))\lambda+\nonumber
\\&&e^{-2Ar^2}r^2(-1+e^{Ar^2}-3Br^2-B^2r^4+Ar^2(2+Br^2))\times\nonumber
\\&&(-e^{Ar^2}(r^2-2\lambda)-2(1+3Br^2+B^2r^4-Ar^2(2+Br^2))\lambda)+e^{-2Ar^2}r^2\times\nonumber
\\&&(-1+e^{Ar^2}+Ar^2-Br^2)(e^{Ar^2}(r^2-4\lambda)+4(1+3Br^2+B^2r^4-Ar^2(2+Br^2))\lambda)\bigg].\label{10}
\end{eqnarray}

\section{Physical Analysis and Graphical Representation}

This section presents the physical properties of the solutions regarding EOS, anisotropic behavior, energy conditions, TOV and matching conditions
along with stability analysis of three different compact stars, Her $X1$, SAX J $1808$-$3658$ and $4$U $1820$-$30$.
Many EOS parameters have been considered in literature in different cosmological contexts.
Staykov et al. \cite{32e} used two different hadronic parameters and a strange matter EOS parameter to study the structure of rotating neutron stars.
Quadratic EOS parameters have been used to study the properties of compact stars \cite{32efg, 32efgh}. For the sake of simplicity, in this work we assume linear EOS \cite{35a}:
\begin{equation}
p_r=\emph{w}_r\rho,~~~~~~~~~~~p_t=\emph{w}_t\rho.
\end{equation}
It is important to note here that the EOS parameters are dependent on radius rather than a constant quantity as in the ordinary matter distribution. The non constant behavior is due to the usual matter and exotic matter contributions.
Figs. $(1)$ and $(2)$ show the variation of EOS parameters $\omega_r$ and $\omega_t$ respectively.
It is obvious from the figures that effective EOS in our model are the same as in
normal matter distribution \cite{35}, i.e.
\begin{equation}
0<\omega_r<1,~~~~0<\omega_t<1.
\end{equation}

\subsection{Anisotropy Measure of the Compact Stars}

The variation of energy density $\rho$, radial pressure $p_r$ and tangential pressure $p_t$ can be observed in
Figs. $(3)$-$(5)$.
The behavior of $\frac{d\rho}{dr}$ and $\frac{dp_r}{dr}$ in Figs. $(6)$ and $(7)$  reveals that $\frac{d\rho}{dr}<0$ and $\frac{dp_r}{dr}<0$. It shows that with the increase in radius of the compact star,  both energy density and radial pressure decrease. We analyze the variation of $\frac{d\rho}{dr}$ and $\frac{dp_r}{dr}$ at the center of the compact star $r=0$ and found that
\begin{eqnarray}\label{11}
\frac{d\rho}{dr}=0,~~~~ \frac{dp_r}{dr}=0,~~~~\frac{d^2\rho}{dr^2}<0,~~~~\frac{d^2p_r}{dr^2}<0.
\end{eqnarray}
Equation (\ref{11}) gives the maximum value of $\rho$ and $p_r$ at center $r=0$.
The anisotropy measurement $\triangle=\frac{2}{r}(p_t-p_r)$ has been shown graphically in Fig. $(14)$. The anisotropy measurement is directed outward when $p_t > p_r$ which results $ \triangle>0$ and directed inward when $p_t < p_r$ which results $\triangle<0$. It is depicted in Fig. $(14)$ that for the large values of $r$, $\triangle>0$ for all stars implying that the anisotropic force permits the construction of great massive configurations. It is worthy to mention that anisotropy measurement vanishes at the center of the star.
\begin{figure}\center
\begin{tabular}{cccc}
\epsfig{file=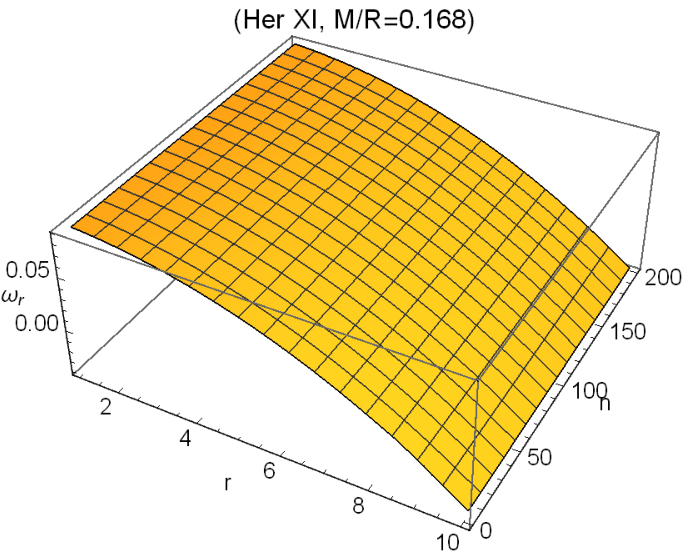,width=0.35\linewidth} &
\epsfig{file=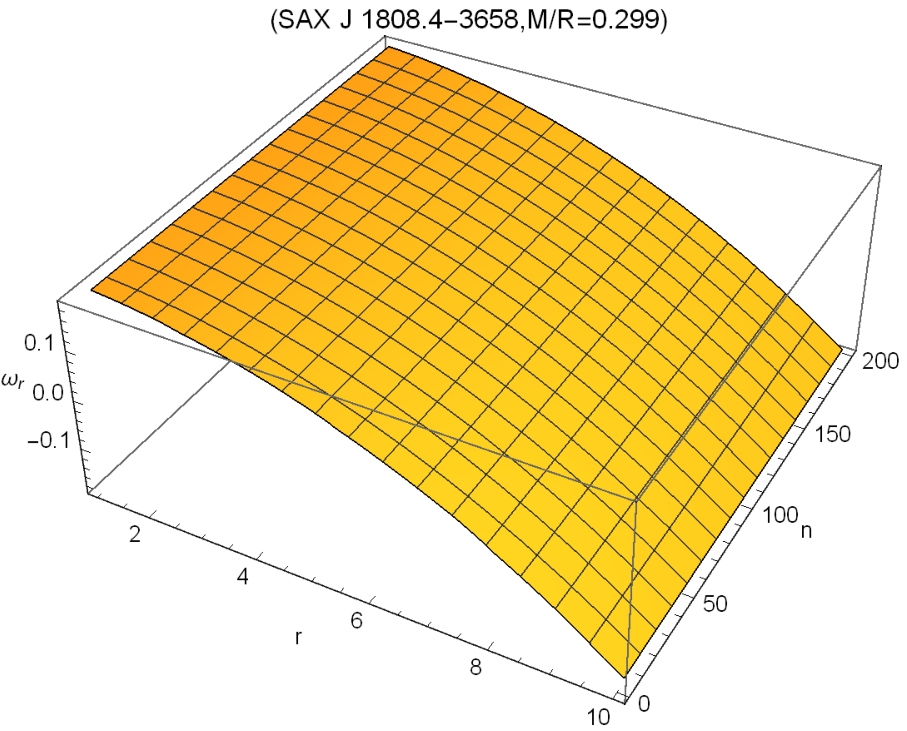,width=0.35\linewidth} &
\epsfig{file=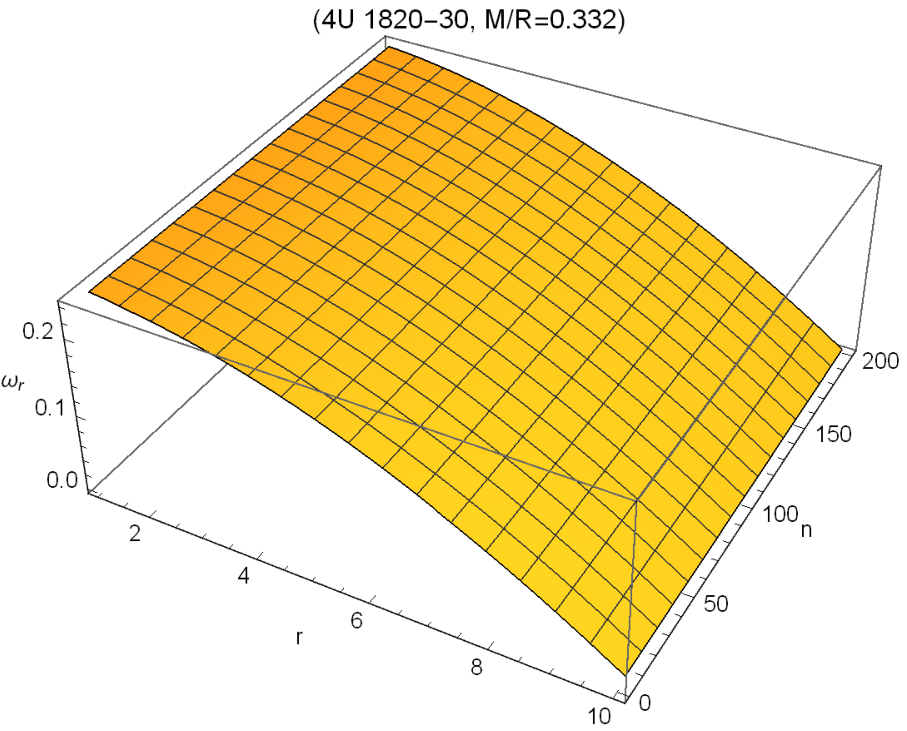,width=0.35\linewidth} \\
\end{tabular}
\caption{Variation of EOS parameter $\omega_r$ with radial coordinate $r$(km) and model parameter $n$}\center
\end{figure}
\begin{figure}\center
\begin{tabular}{cccc}
\epsfig{file=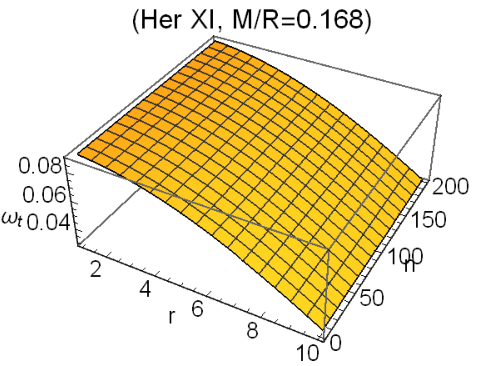,width=0.35\linewidth} &
\epsfig{file=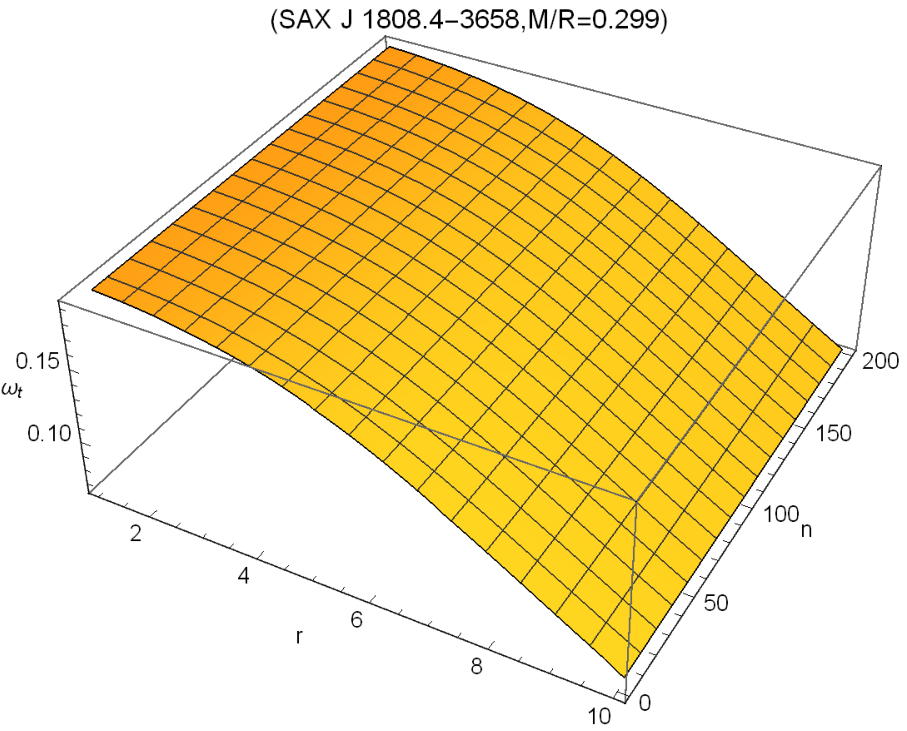,width=0.35\linewidth} &
\epsfig{file=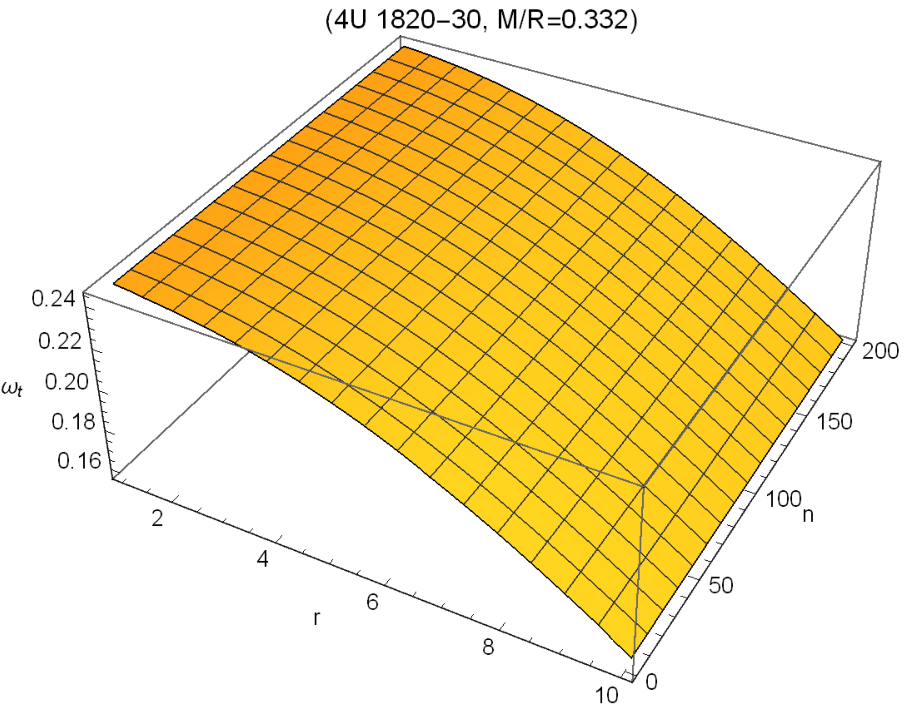,width=0.35\linewidth} \\
\end{tabular}
\caption{Variation of  EOS parameter $\omega_t$ with radial coordinate $r$(km) and model parameter $n$}\center
\end{figure}
\begin{figure}\center
\begin{tabular}{cccc}
\epsfig{file=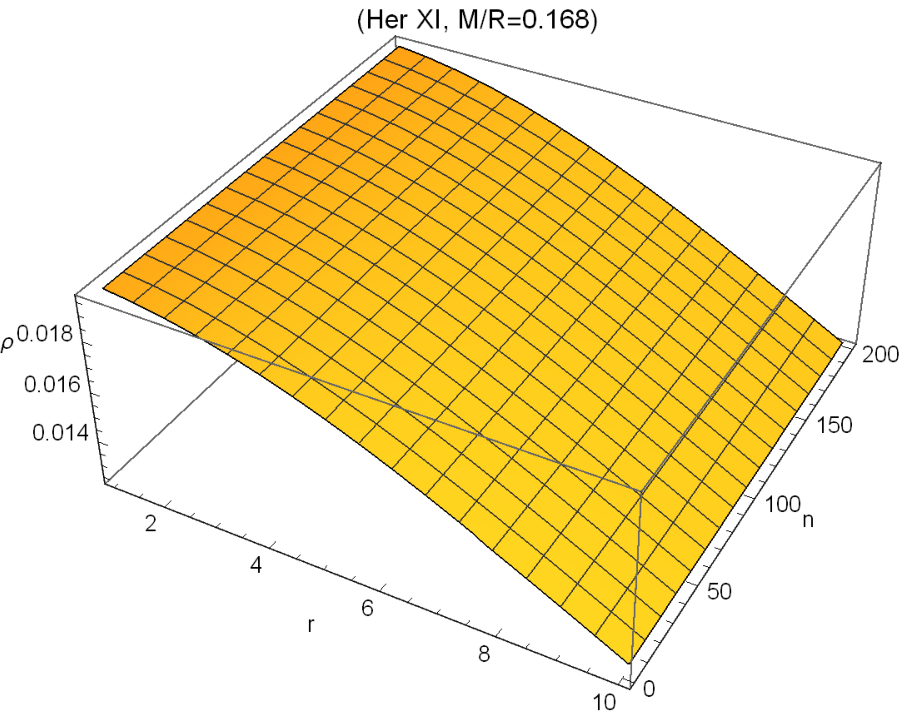,width=0.35\linewidth} &
\epsfig{file=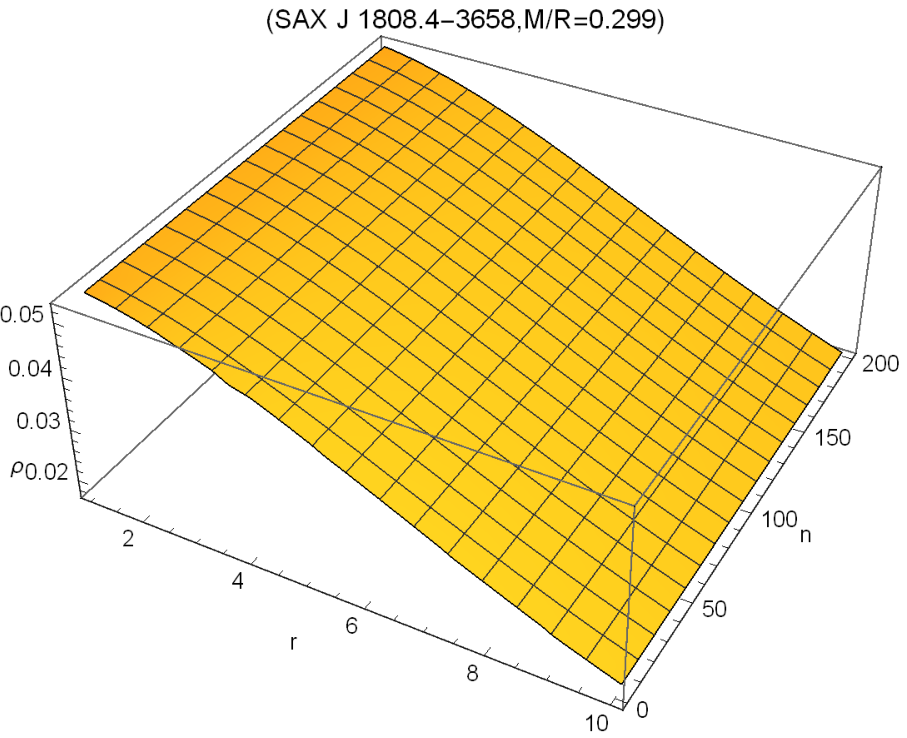,width=0.35\linewidth} &
\epsfig{file=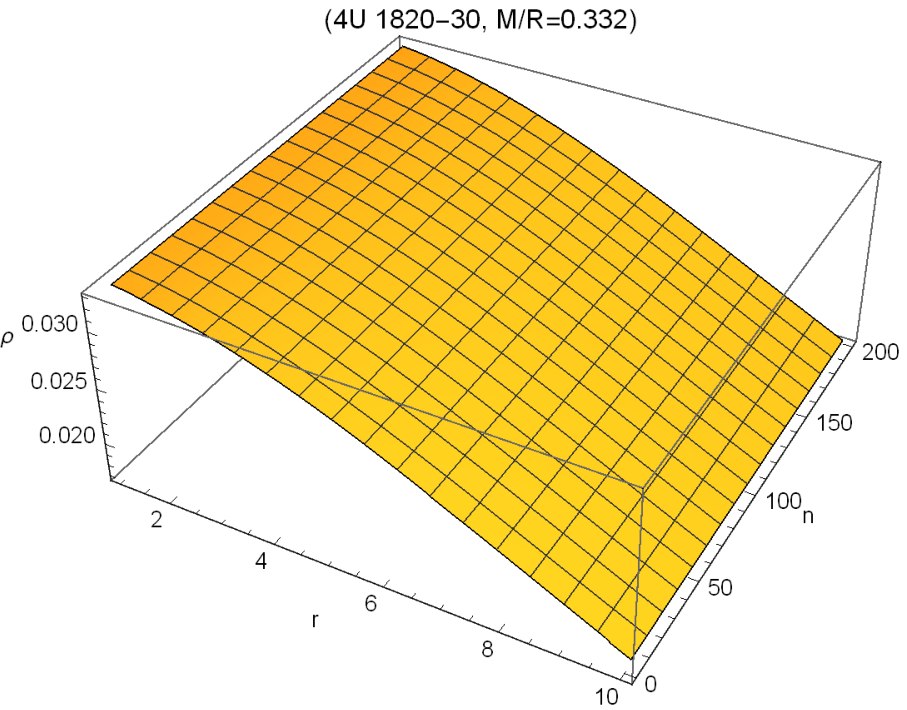,width=0.35\linewidth} \\
\end{tabular}
\caption{Behavior of energy density $\rho$ with radial coordinate $r$(km) and model parameter $n$}\center
\end{figure}
\begin{figure}\center
\begin{tabular}{cccc}
\epsfig{file=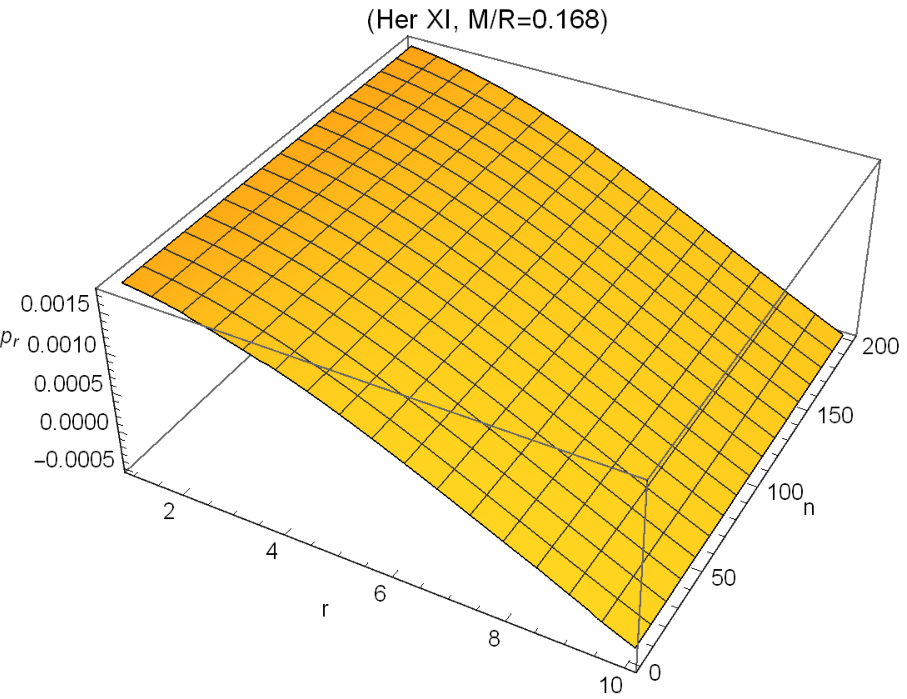,width=0.35\linewidth} &
\epsfig{file=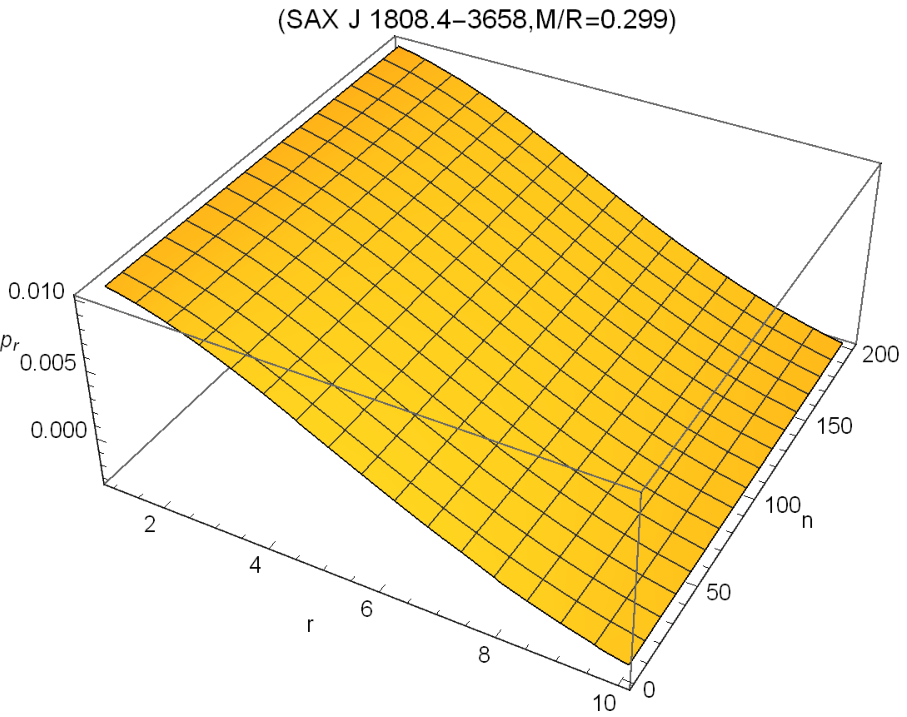,width=0.35\linewidth} &
\epsfig{file=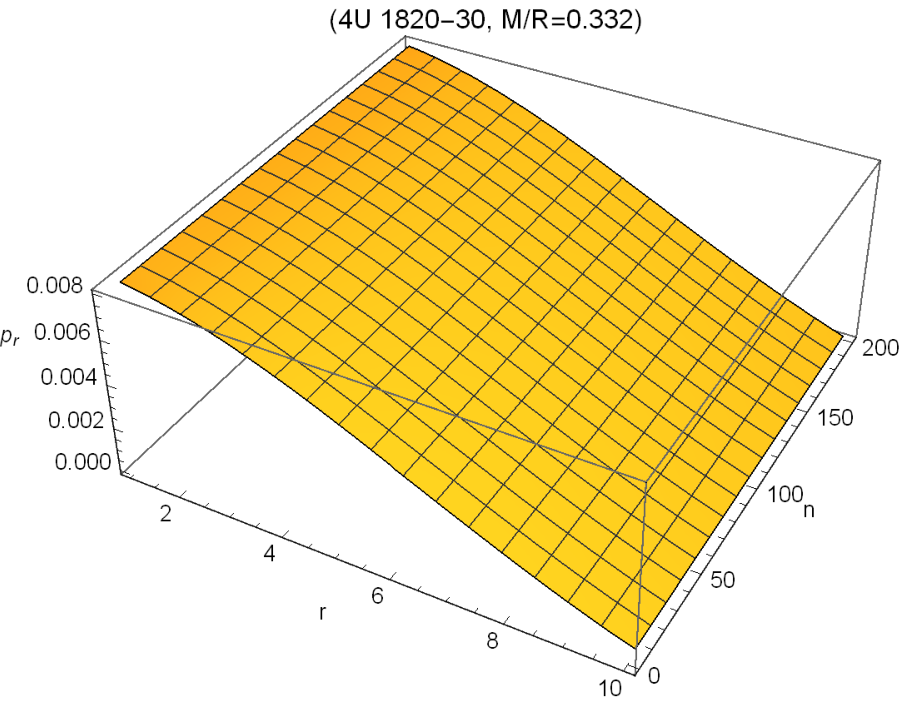,width=0.35\linewidth} \\
\end{tabular}
\caption{Behavior of radial pressure $p_r$ with radial coordinate $r$(km) and model parameter $n$}\center
\end{figure}
\begin{figure}\center
\begin{tabular}{cccc}
\epsfig{file=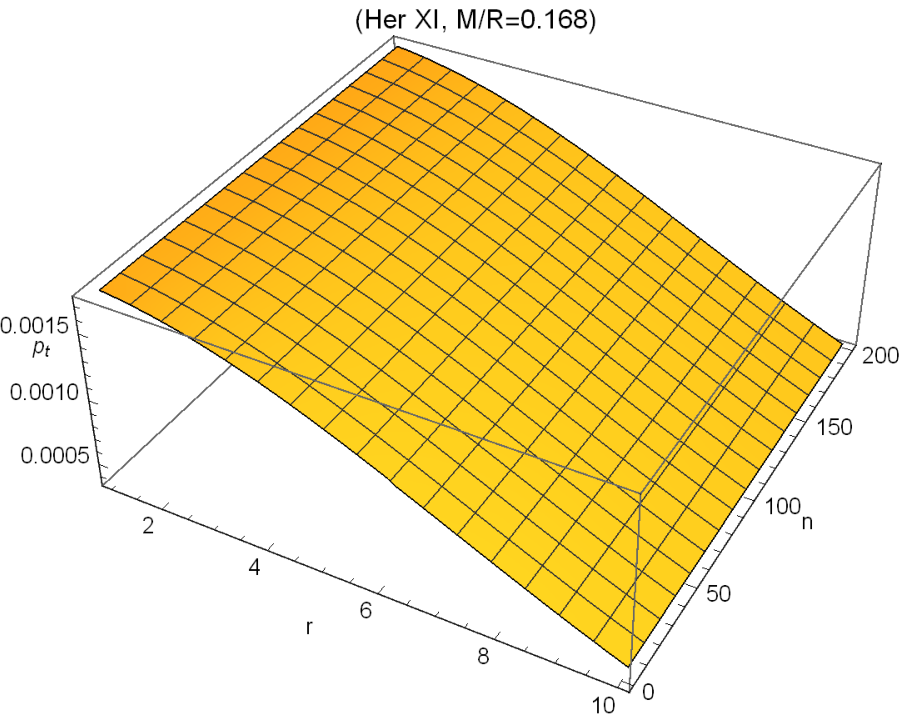,width=0.35\linewidth} &
\epsfig{file=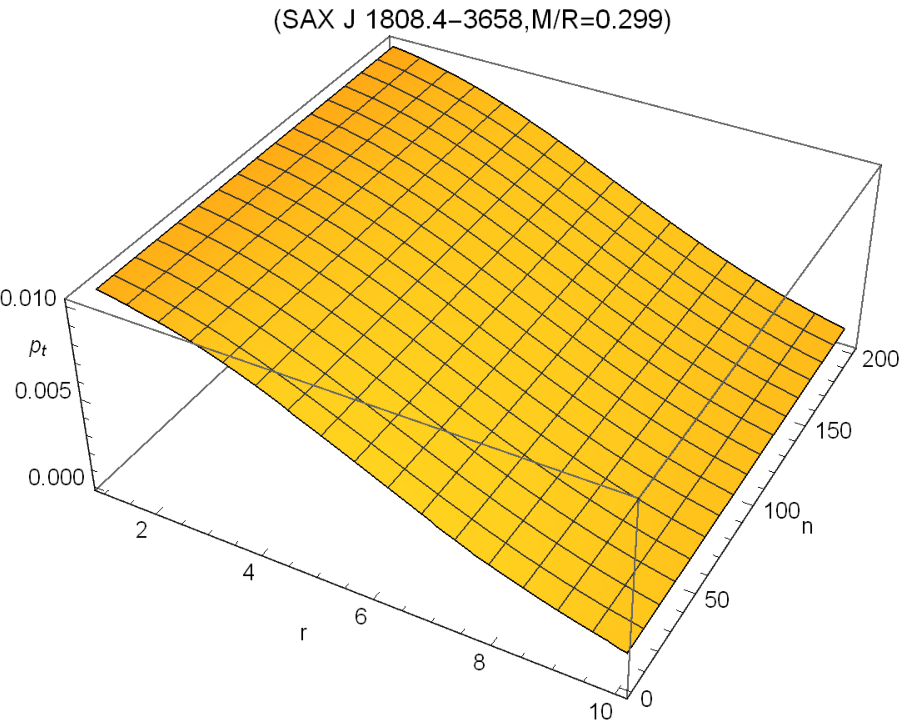,width=0.35\linewidth} &
\epsfig{file=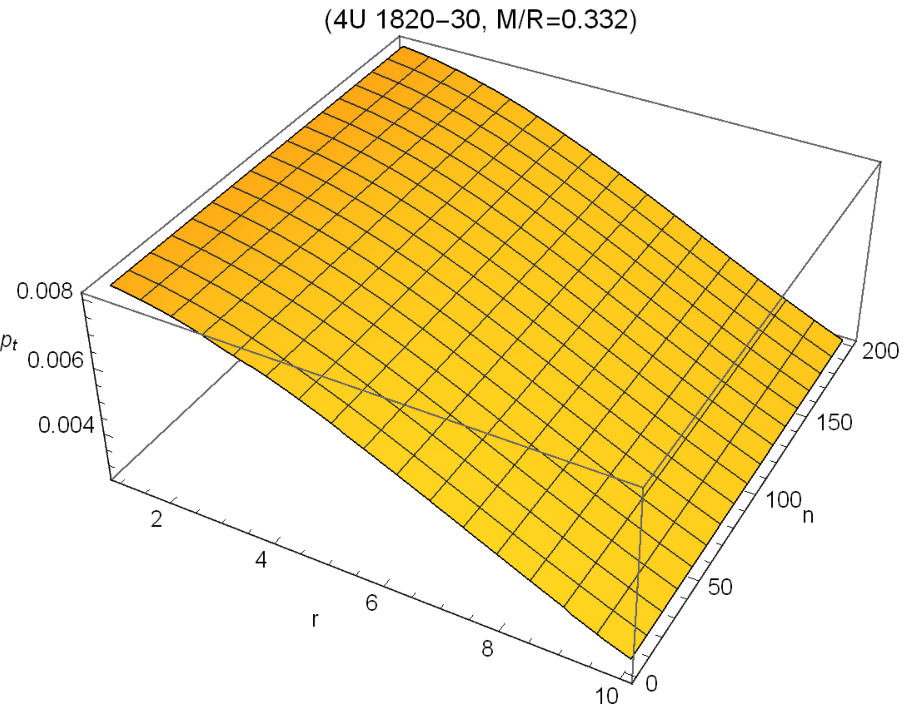,width=0.35\linewidth} \\
\end{tabular}
\caption{Behavior of transverse pressure  $p_t$ with radial coordinate $r$(km) and model parameter $n$}\center
\end{figure}
\begin{figure}\center
\begin{tabular}{cccc}
\epsfig{file=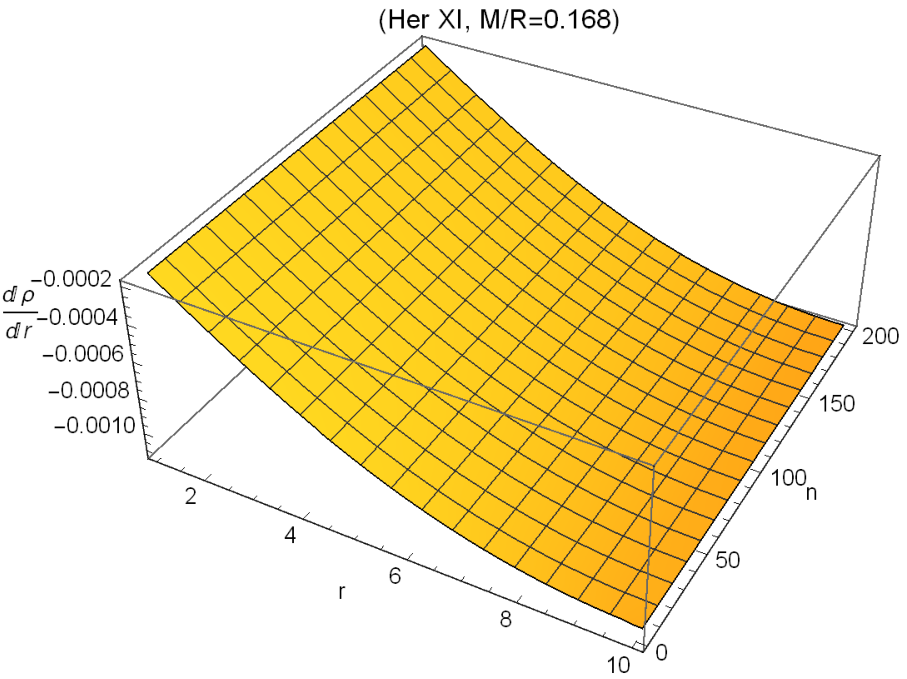,width=0.35\linewidth} &
\epsfig{file=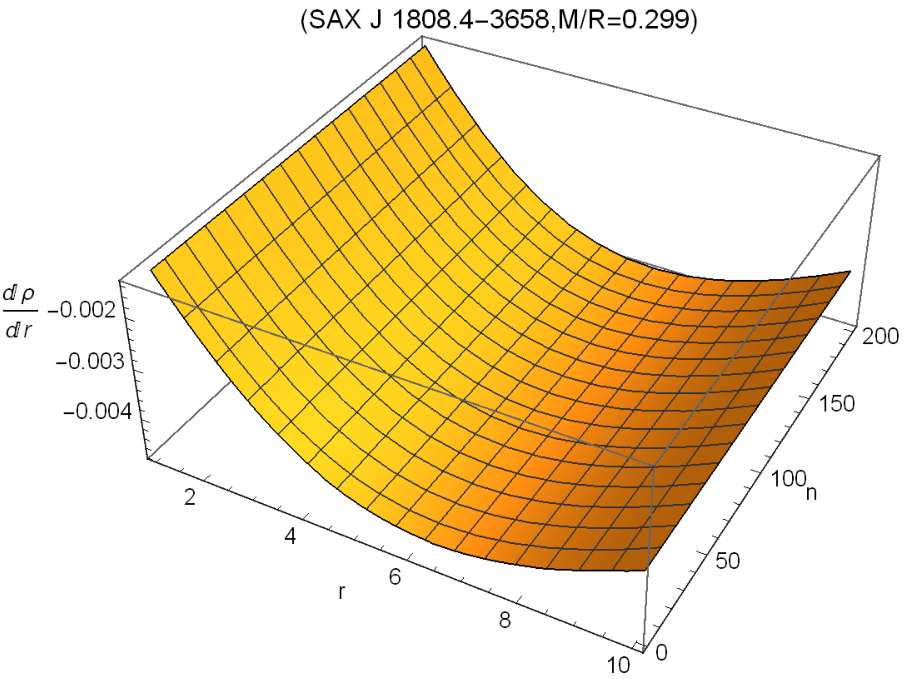,width=0.35\linewidth} &
\epsfig{file=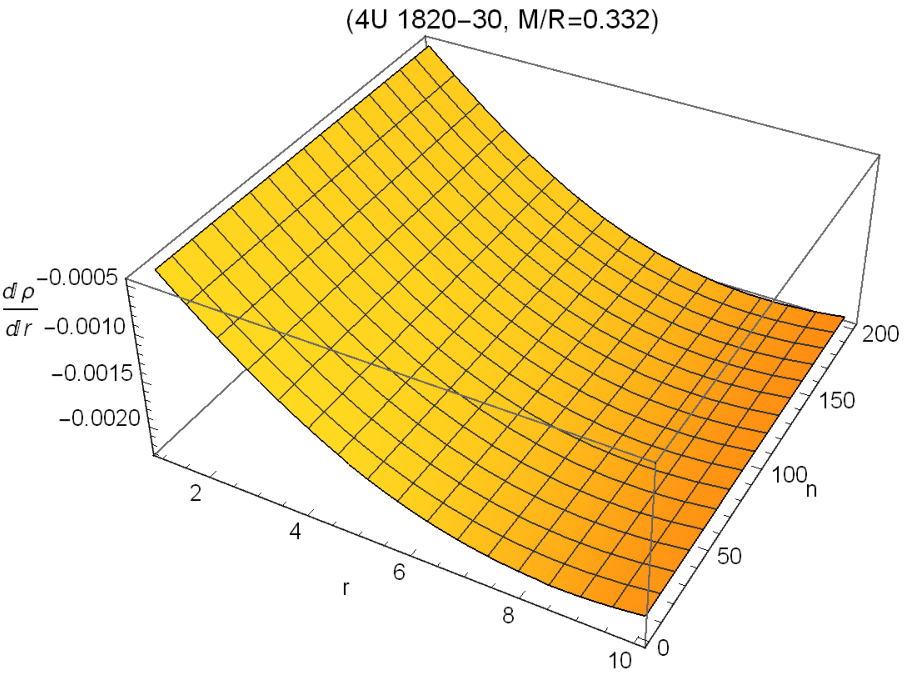,width=0.35\linewidth} \\
\end{tabular}
\caption{Behavior of  $\frac{d\rho}{dr}$ with radial coordinate $r$(km) and model parameter $n$}\center
\end{figure}
\begin{figure}\center
\begin{tabular}{cccc}
\epsfig{file=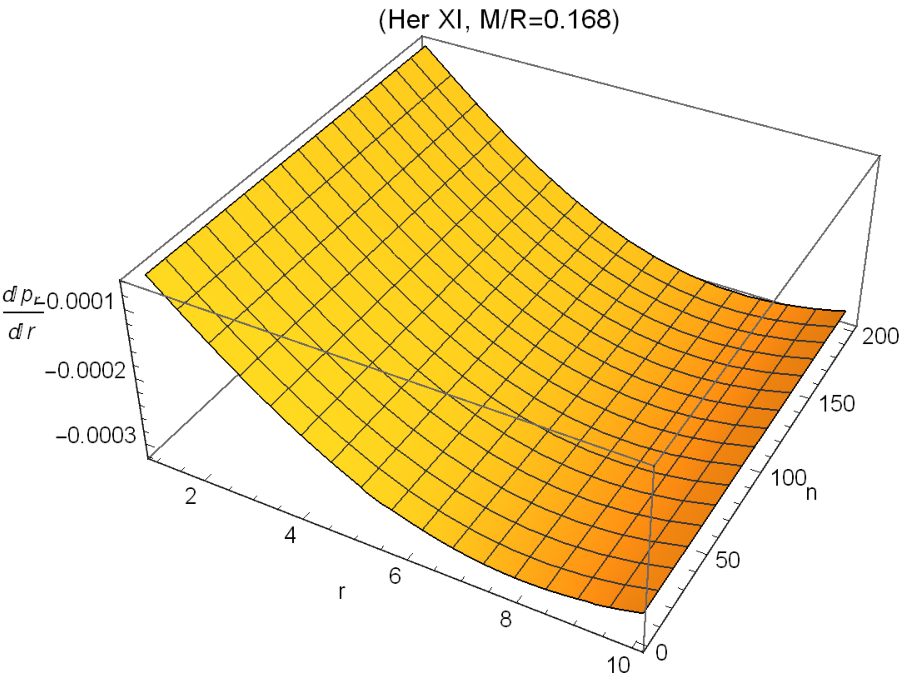,width=0.35\linewidth} &
\epsfig{file=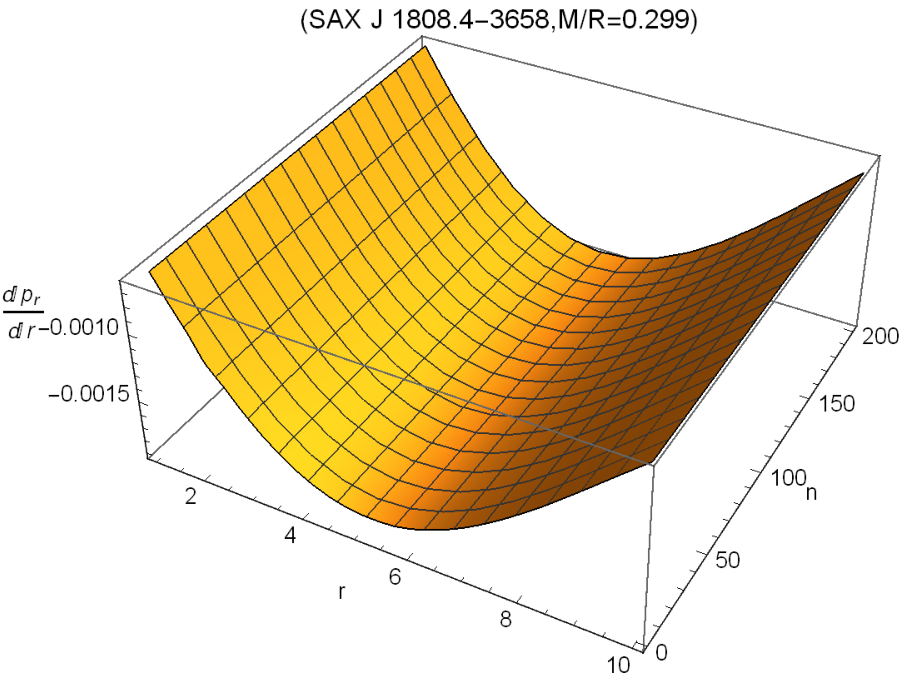,width=0.35\linewidth} &
\epsfig{file=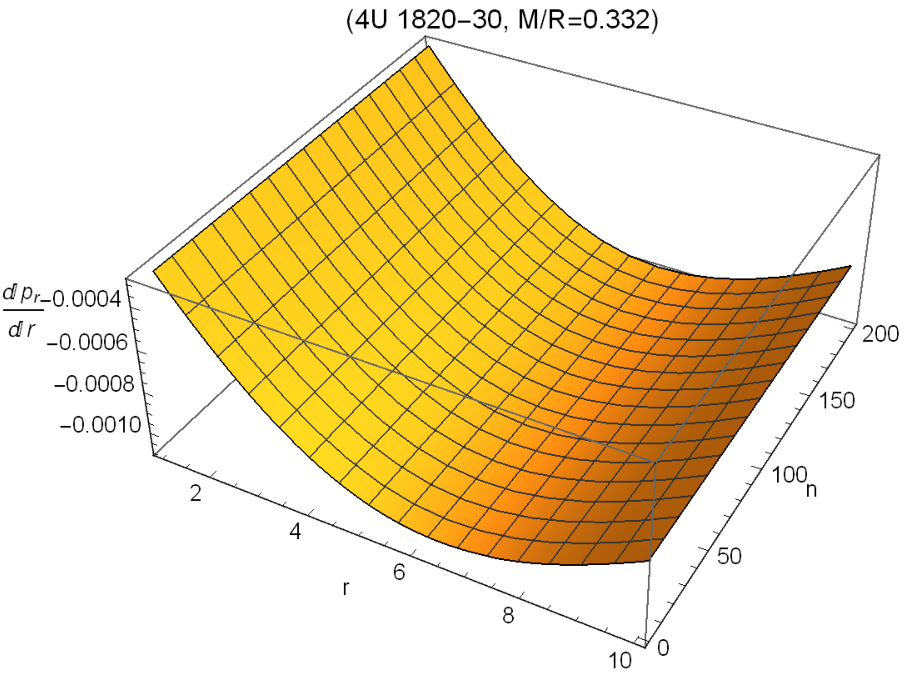,width=0.35\linewidth} \\
\end{tabular}
\caption{Behavior of  $\frac{dp_r}{dr}$ with radial coordinate $r$(km) and model parameter $n$}\center
\end{figure}

\subsection{The Matching Conditions with Schwarzchild Exterior metric}

The interior metric of the boundary surface  remains same in the interior and exterior geometry of the compact star. It justifies the continuity of the metric components for the boundary surface of the star. Many choices for the matching conditions are possible and one can consider the vacuum outside a general spherically symmetric space time with suitable boundary conditions \cite{27}. For the present analysis, we consider Schwarzchild solution for describing the exterior geometry.  Many authors have considered Schwarzchild solution for this purpose giving some interesting results \cite{27a}-\cite{31b}. Therefore, the exterior metric given by Schwarzchild is
\begin{equation}
ds^2=(1-\frac{2M}{r})dt^2-(1-\frac{2M}{r})^{-1}dr^2-r^2(d\theta^2+\sin^2d\varphi^2).
\end{equation}
The intrinsic metric (\ref{3}) for the smooth match at the boundary surface $r=R$ with Schwarzchild exterior metric produces,
\begin{equation}
g^-_{tt}=g^+_{tt},~~~g^-_{rr}=g^+_{rr},~~~\frac{\partial g^-_{tt}}{\partial r}=\frac{\partial g^+_{tt}}{\partial r},
\end{equation}
where interior solutions and exterior solutions are represented by (-) and (+). By the matching of interior and exterior metrics, we obtain
\begin{eqnarray}\nonumber
\\&&A=-\frac{1}{R^2}\ln(1-\frac{2M}{R}),
\\&&B=\frac{M}{R^3}(1-\frac{2M}{R})^{-1},
\\&&C=\ln(1-\frac{2M}{R})-\frac{M}{R}(1-\frac{2M}{R})^{-1}.
\end{eqnarray}
Using the approximate values of $M$ and $R$ for the compact stars under observation, the constants $A$ and $B$ are given in the following table \cite{35}
\begin{center}
\begin{tabular}{ |c|c|c|c|c|c|c| }
 \hline
 Models  &      M           & R(km)  & $\alpha=\frac{M}{R}$& $A(km^{-2})$     & $B(km^{-2})$   & $Z_{s}$ \\
 \hline
 \textbf{Her XI } & 0.88$M_{\odot}$  & 7.7    &  0.168              & 0.0069062764281   & 0.0042673646183 & 0.23  \\
 \hline
 \textbf{SAX J}  & 1.435$M_{\odot}$ & 7.07   &  0.299              & 0.018231569740    & 0.014880115692  & 0.57  \\
 \hline
 \textbf{4U}  & 2.25$M_{\odot}$  & 10.0   &  0.332              & 0.010906441192    & 0.0098809523811 & 0.073 \\
 \hline
\end{tabular}
\end{center}

\subsection{The Energy Bounds}

The energy bounds have gained much importance in the discussion of some important issues in cosmology.
In fact, one can investigate the validity of second law of black hole thermodynamics and Hawking-Penrose singularity theorems using energy conditions \cite{Hawking}. Many interesting results have been reported in cosmology using the energy bounds \cite{Santos}-\cite{Bertolami}.
These energy conditions are defined as
\begin{eqnarray}\nonumber
\\&&NEC:~~~~~~\rho+\emph{p}_r\geq0,~~~\rho+\emph{p}_t\geq0\nonumber,
\\&&WEC:~~~~~~\rho\geq0,~~~\rho+\emph{p}_r\geq0,~~~\rho+\emph{p}_t\geq0\nonumber,
\\&&SEC:~~~~~~\rho+\emph{p}_r\geq0,~~~\rho+\emph{p}_t\geq0,~~~\rho+\emph{p}_r+2\emph{p}_t\geq0\nonumber,
\\&&DEC:~~~~~~\rho>\mid\emph{p}_r\mid,~~~\rho>\mid\emph{p}_t\mid.\nonumber
\end{eqnarray}
where the null energy conditions, weak energy conditions, strong energy conditions and dominant energy conditions are denoted by $NEC$, $WEC$, $SEC$  and $DEC$ respectively.
In Fig. $(8)$, it is evident that all energy conditions are satisfied for Her $X1$. The energy conditions are also satisfied for other two stars but the graphs are not shown here.
\begin{figure}\center
\begin{tabular}{cccc}
\epsfig{file=rho.HerX1.eps,width=0.35\linewidth} &
\epsfig{file=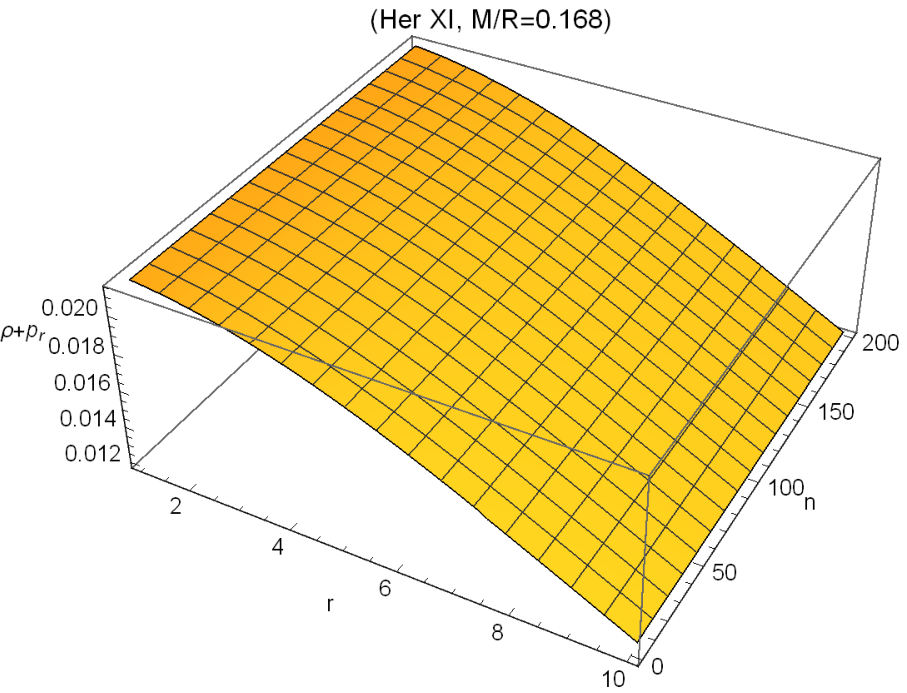,width=0.35\linewidth} &\\
\epsfig{file=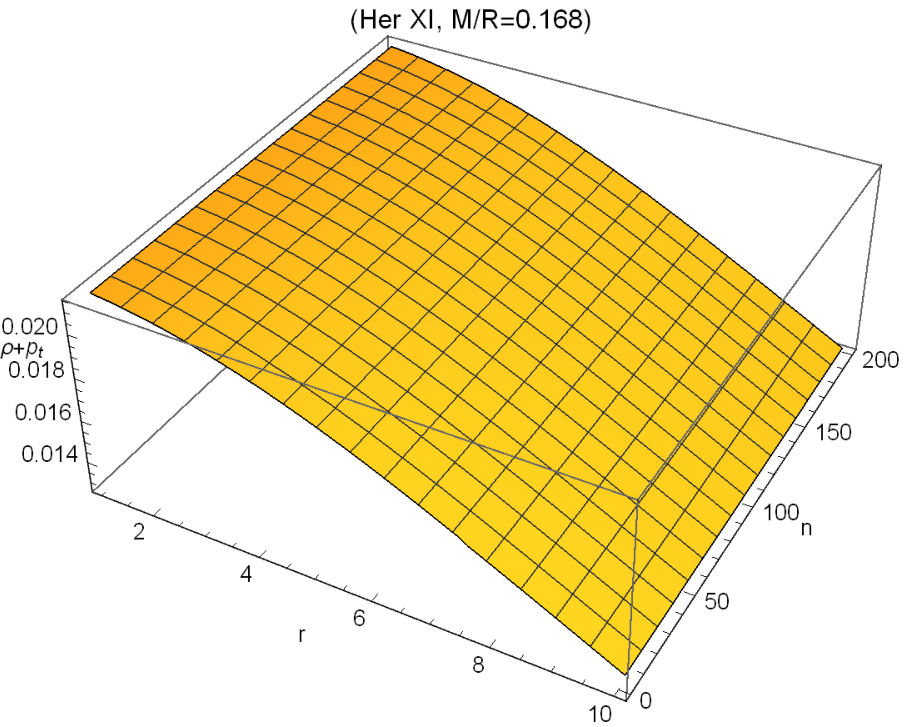,width=0.35\linewidth} &
\epsfig{file=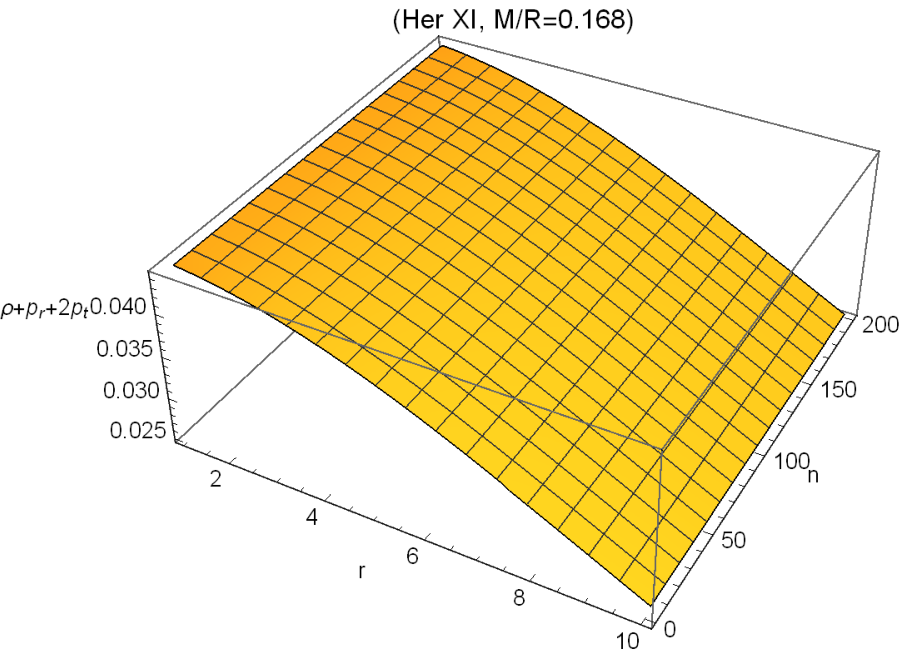,width=0.35\linewidth} &\\
\epsfig{file=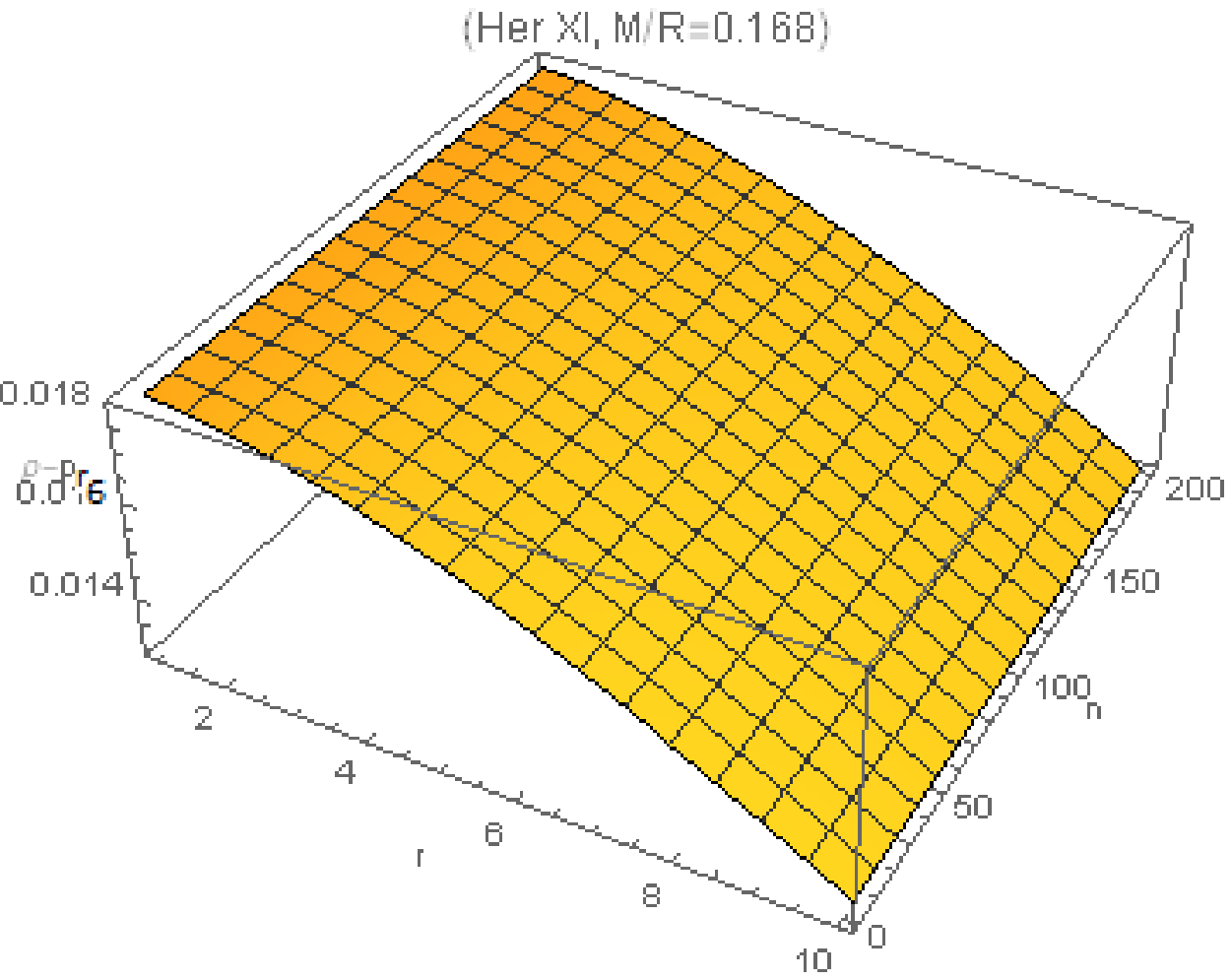,width=0.35\linewidth} &
\epsfig{file=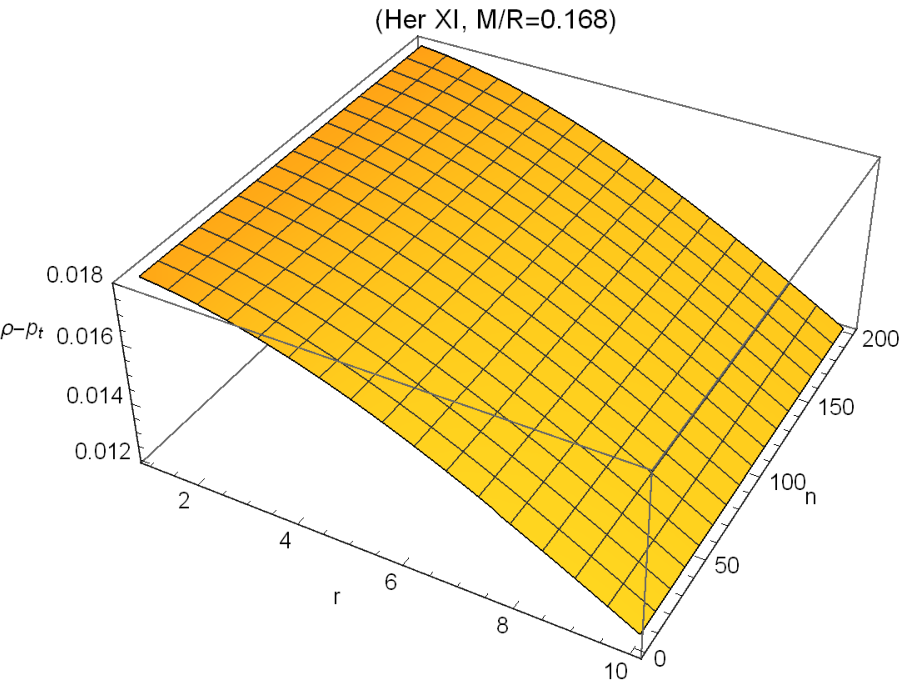,width=0.35\linewidth} &\\
\end{tabular}
\caption{Energy Conditions for compact star Her $X1$ with radial coordinate $r$(km) and model parameter $n$}\center
\end{figure}

\subsection{Implementation of Tolman-Oppenheimer-Volkoff Equation}

The TOV equation is expressed in the following generalized form
\begin{equation}\label{12}
\frac{dp_r}{dr}+\frac{\nu\acute{} (\rho+p_r)}{2}+\frac{2(p_r-p_t)}{r}=0.
\end{equation}
It follows from equation (\ref{12}),
\begin{eqnarray}
F_g+F_h+F_a=0,
\end{eqnarray}
where
\begin{equation}
F_g=-Br(\rho+p_r),~~~F_h=-\frac{dp_r}{dr},~~~F_a=\frac{2(p_t-p_r)}{r}.
\end{equation}
Here $F_g$, $F_h$ and $F_a$ are gravitating force, hydrostatic force and anisotropic pressure force for compact stars. Using the values of $\rho$, $p_r$ and $p_t$ from equations (\ref{8})-(\ref{10}), for compact star $4U$, Fig.$(9)$ shows the behavior of these forces. Fig. $(10)$ depicts that the TOV equation is satisfied for compact star $4$U. The TOV equation is satisfied for the other two stars as well and graphical analysis is not presented here.
\begin{figure}\center
\begin{tabular}{cccc}
\epsfig{file=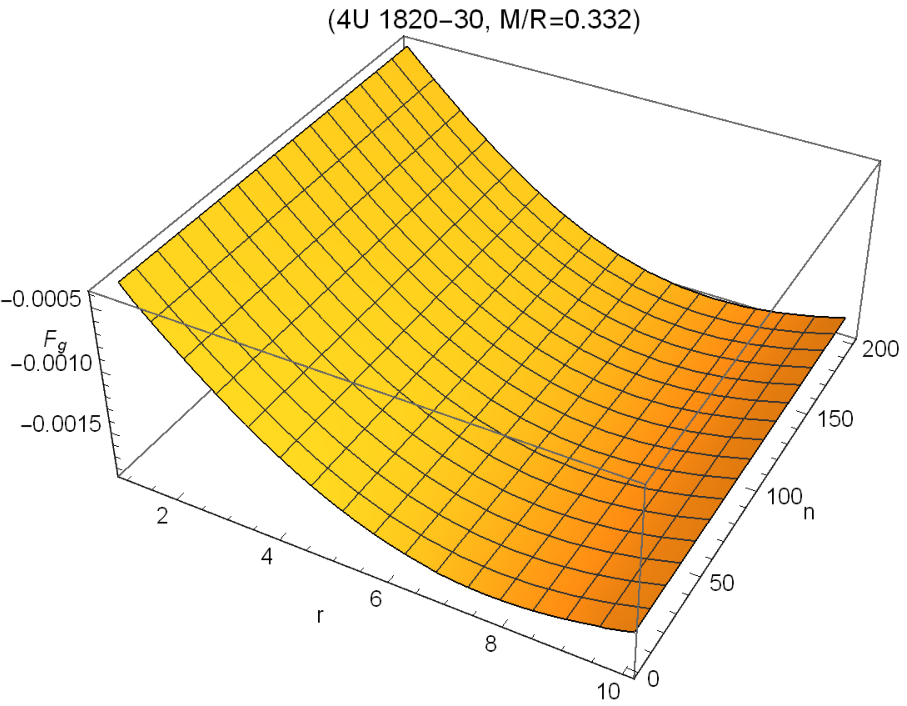,width=0.35\linewidth} &
\epsfig{file=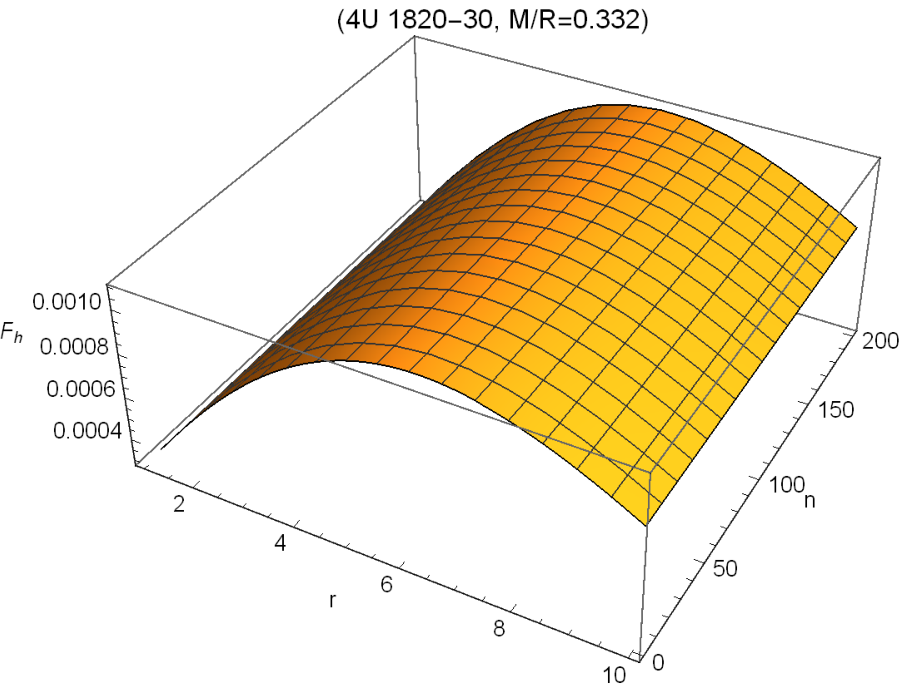,width=0.35\linewidth} &
\epsfig{file=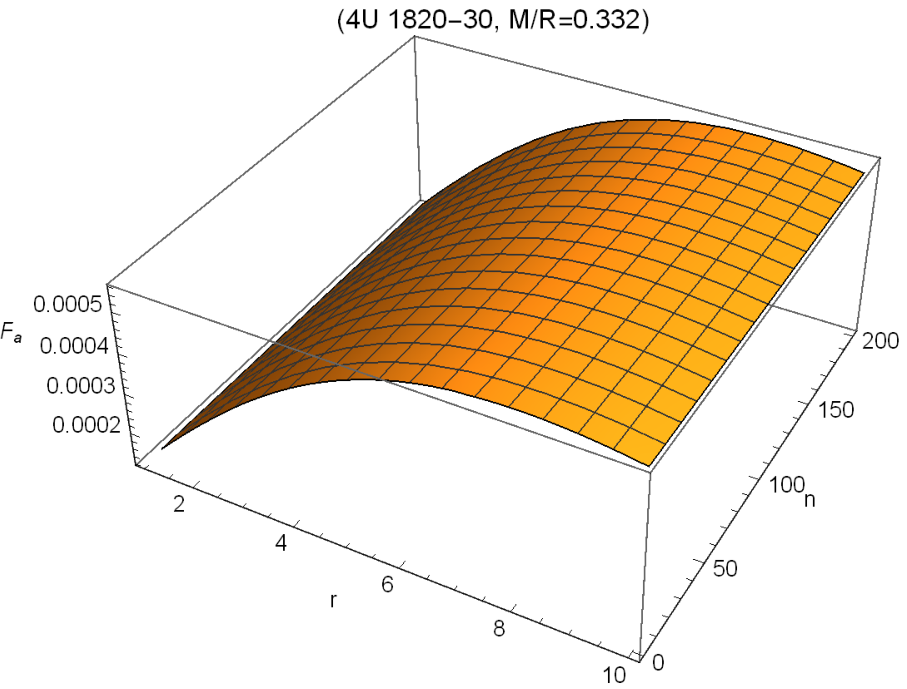,width=0.35\linewidth} \\
\end{tabular}
\caption{Behavior of three different forces namely, gravitating, hydrostatic and pressure anisotropic forces in compact star candidate $4U$}\center
\end{figure}
\begin{figure}\center
\begin{tabular}{cccc}
\epsfig{file=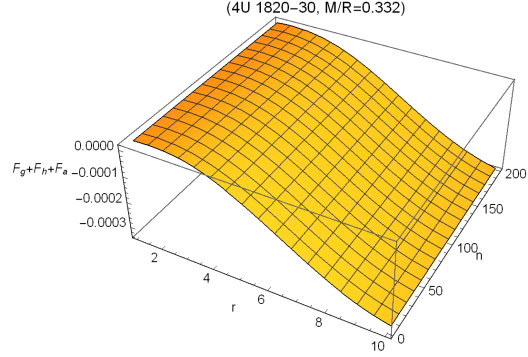,width=0.50\linewidth} \\
\end{tabular}
\caption{Plot of TOV equation for compact star candidate $4U$}\center
\end{figure}

\subsection{Stability Analysis}

Now we find the radial sound speed $\upsilon_{sr}$ and transverse sound speed $\upsilon_{st}$ to determine the stability of our model where
\begin{eqnarray}
\upsilon_{sr}=\frac{dp_r}{d\rho},~~~~~~ \upsilon_{st}=\frac{dp_t}{d\rho}.
\end{eqnarray}
For a stable model following conditions must hold.
\begin{eqnarray}\label{13}
0\leq\upsilon_{sr}^2\leq1,~~~~~~0\leq\upsilon_{st}^2\leq1.
\end{eqnarray}
The conditions (\ref{13}) are exhibited graphically in Figs. ($11$) and ($12$) and show that these conditions are true for considered compact stars. In $1992$, Herrera gave the important concept that for a potentially stable region $\upsilon_{sr}$ is greater than $\upsilon_{st}$ \cite{37}. Therefore, behavior of $\upsilon_{sr}^2-\upsilon_{st}^2$ is shown in Fig.($13$). It can be observed that $\mid \upsilon_{sr}^2-\upsilon_{st}^2\mid <1$. So, proposed models of compact stars are stable (Fig. $13$).
\begin{figure}\center
\begin{tabular}{cccc}
\epsfig{file=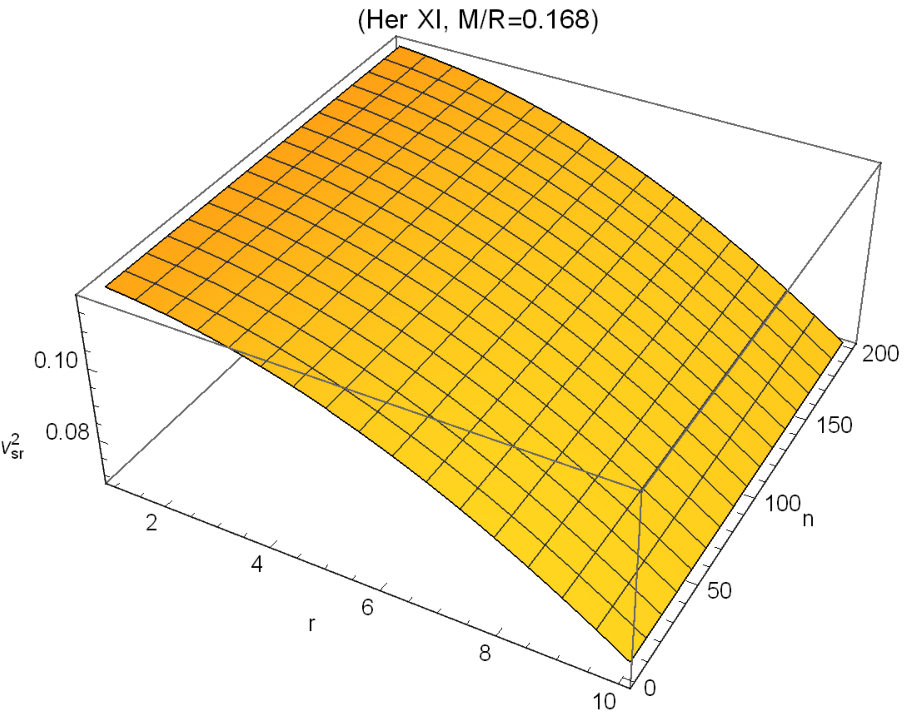,width=0.35\linewidth} &
\epsfig{file=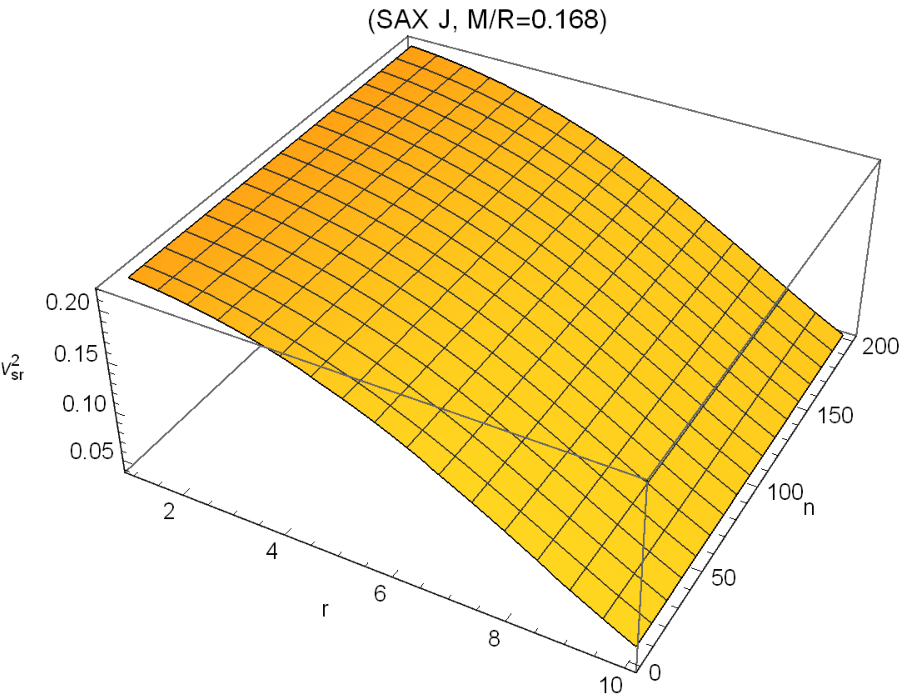,width=0.35\linewidth} &
\epsfig{file=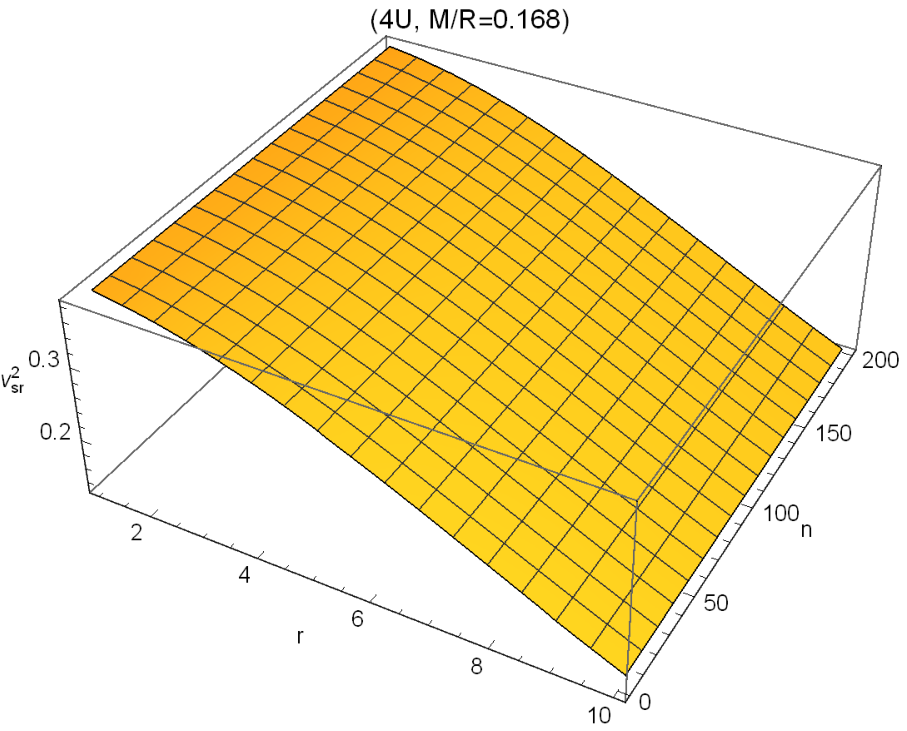,width=0.35\linewidth} \\
\end{tabular}
\caption{Behavior of  $\upsilon^2_{sr}$ with radial coordinate $r$(km) and model parameter $n$ in different compact stars}\center
\end{figure}
\begin{figure}\center
\begin{tabular}{cccc}
\epsfig{file=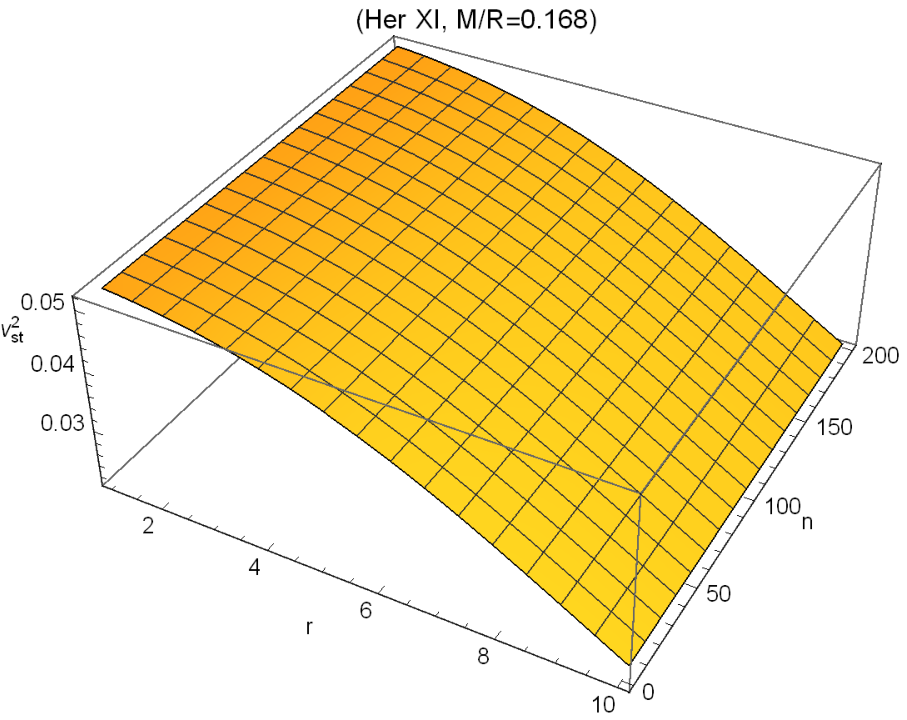,width=0.35\linewidth} &
\epsfig{file=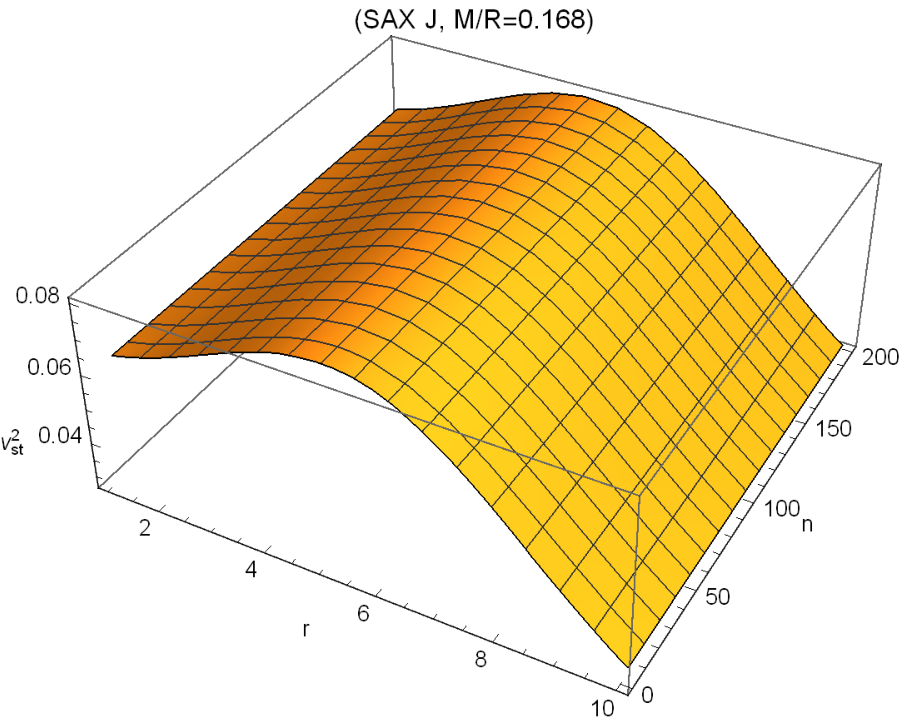,width=0.35\linewidth} &
\epsfig{file=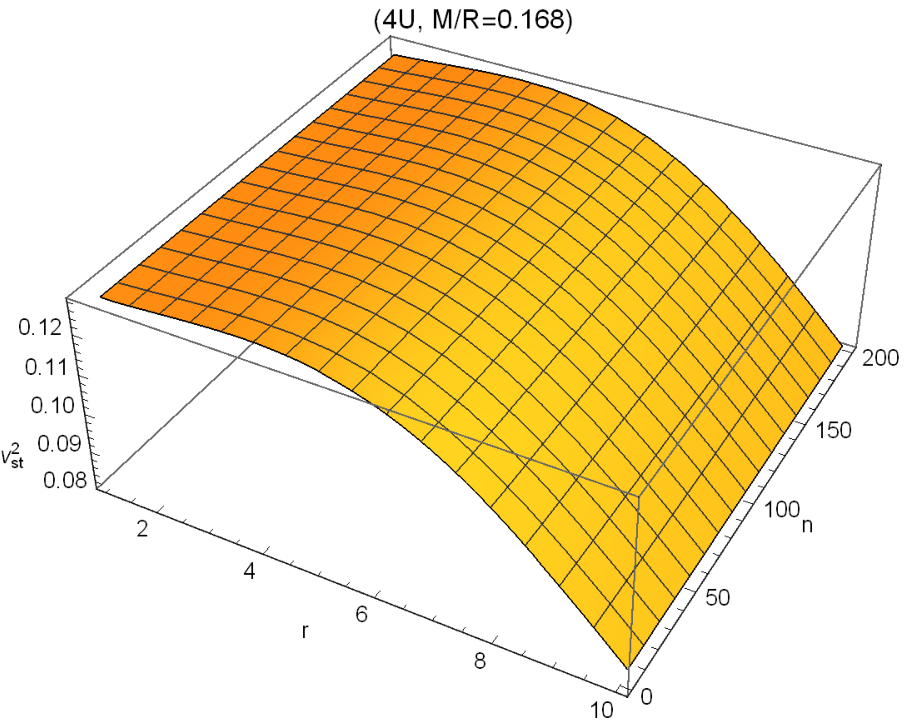,width=0.35\linewidth} \\
\end{tabular}
\caption{Behavior of  $\upsilon^2_{st}$ with radial coordinate $r$(km) and model parameter $n$ in different compact stars}\center
\end{figure}
\begin{figure}\center
\begin{tabular}{cccc}
\epsfig{file=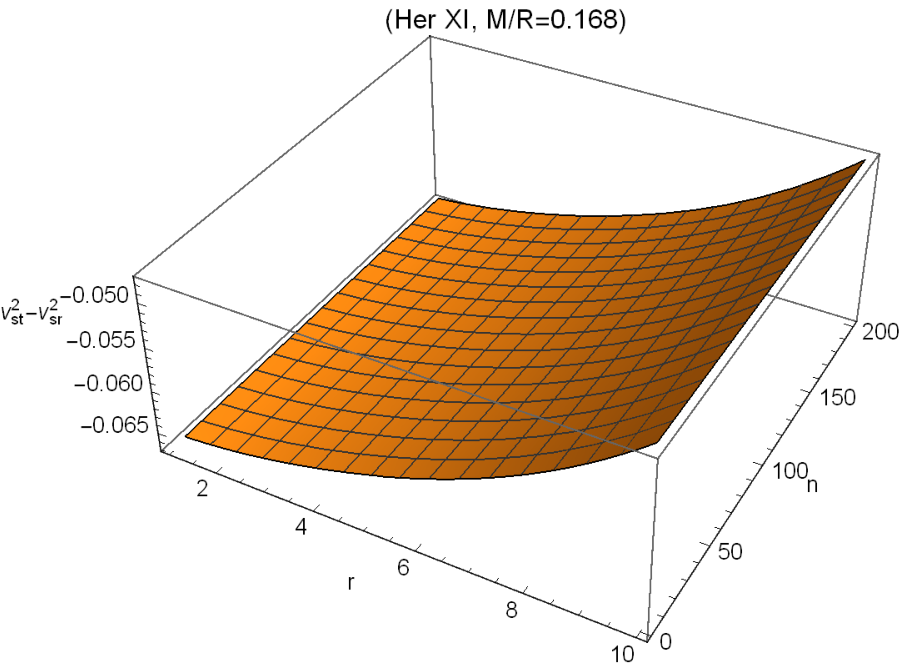,width=0.35\linewidth} &
\epsfig{file=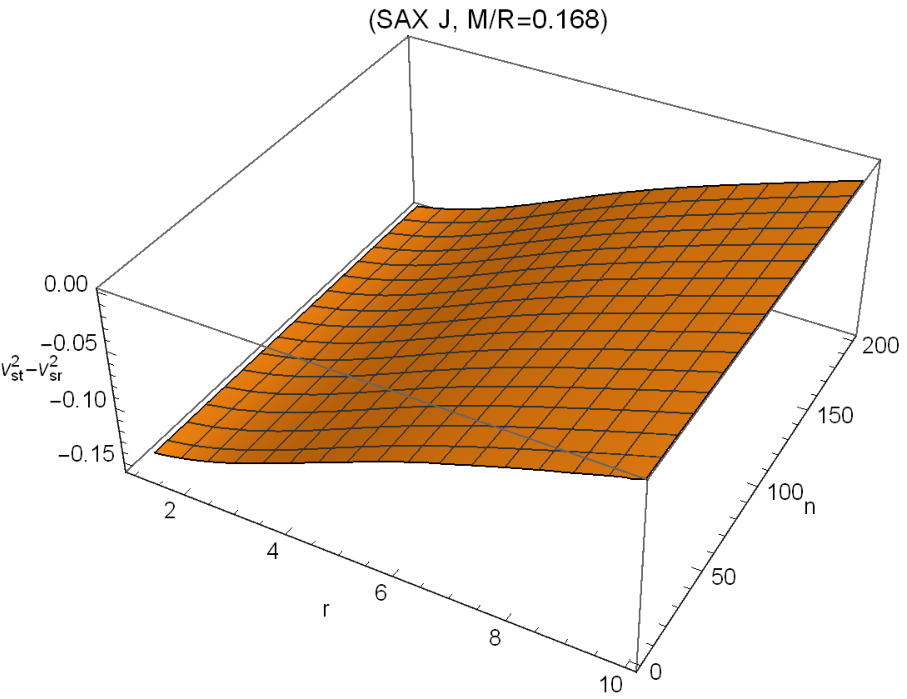,width=0.35\linewidth} &
\epsfig{file=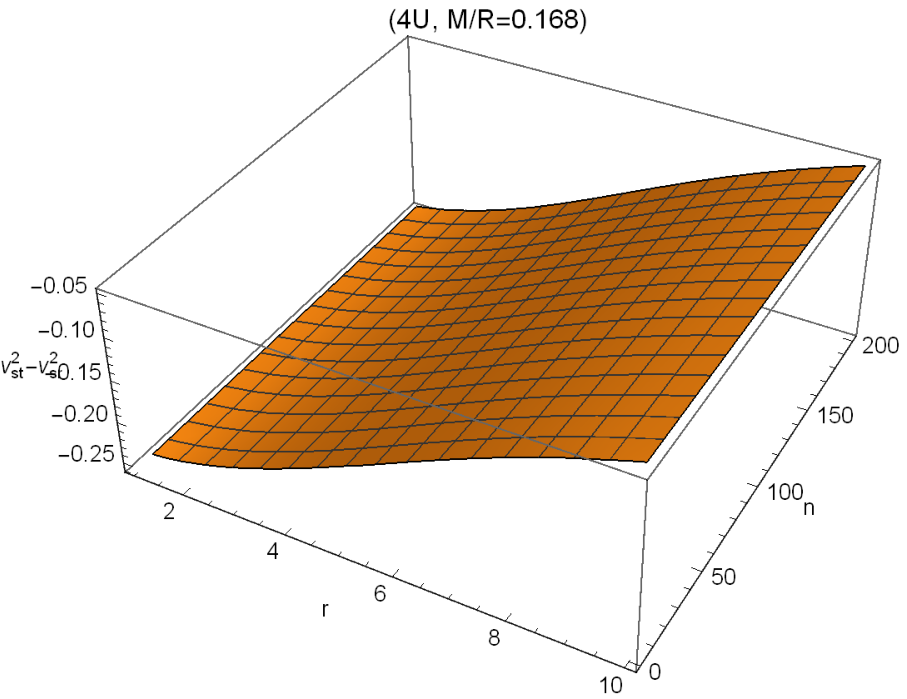,width=0.35\linewidth} \\
\end{tabular}
\caption{Behavior of  $\upsilon^2_{st}-\upsilon^2_{sr}$ with radial coordinate $r$(km) and model parameter $n$ in different compact stars}\center
\end{figure}
\begin{figure}\center
\begin{tabular}{cccc}
\epsfig{file=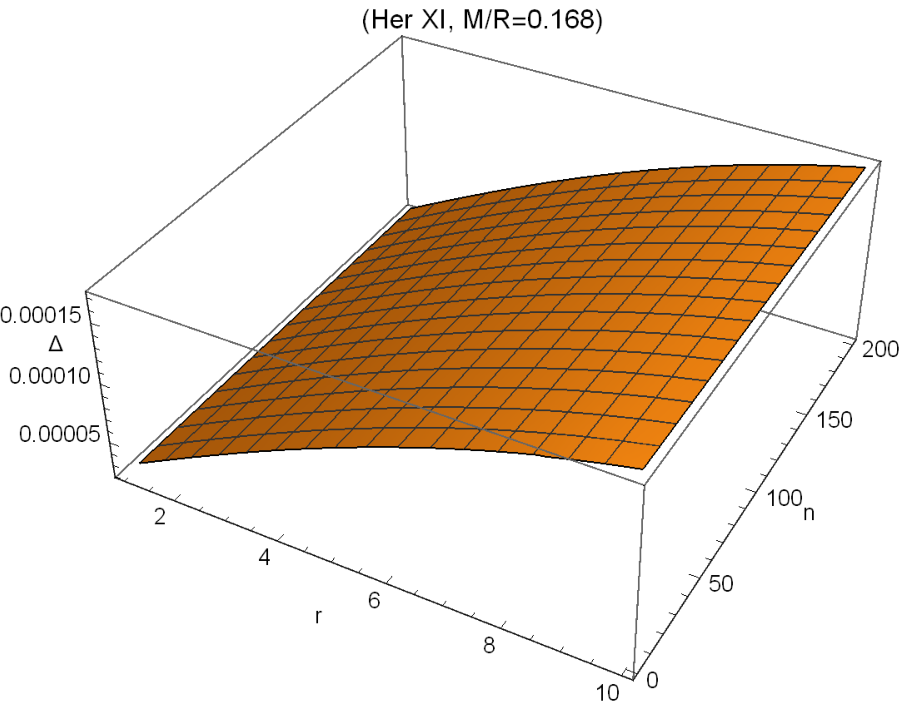,width=0.35\linewidth} &
\epsfig{file=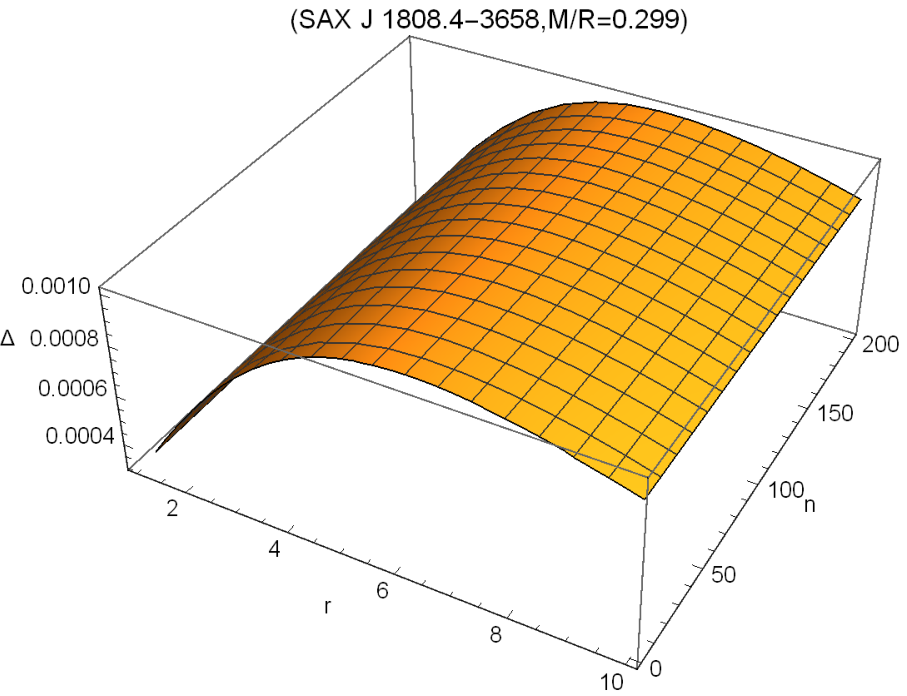,width=0.35\linewidth} &
\epsfig{file=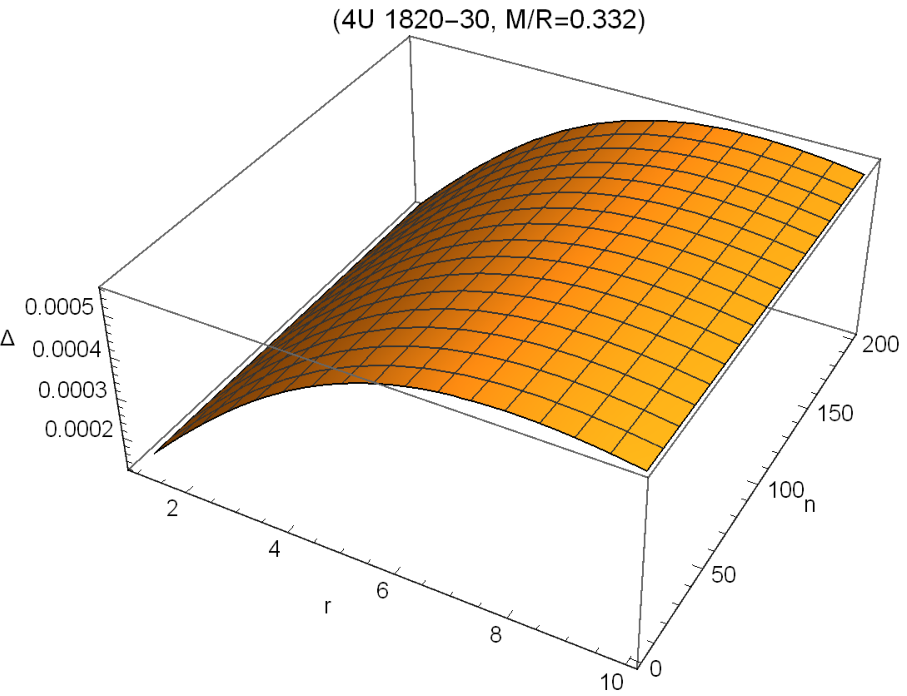,width=0.35\linewidth} \\
\end{tabular}
\caption{variation of  $\triangle$ with radial coordinate $r$(km) and model parameter $n$ in different compact stars}\center
\end{figure}

\section{Final Remarks}

The current study deals with the physical attributes of the compact stars in the scenario of $f(R,G)$ gravity. In this regard, interior solutions are figured out for three compact stars namely, Her $X1$, SAX J $1808$-$3658$ and 4U $1820$-$30$, with the assumption that they have anisotropic internal structure.  The analytic solutions of the interior metric in $f(R,G)$ are smoothly matched with Schwarzchild exterior metric. This matching has provided the values of unknown constants $A$, $B$, and $C$ in the form of observed values of radii and masses of the model compact stars. The nature of the stars has been discussed using values of these constants.  The analysis of the physical attributes leads one to conclude the following results as regards the anisotropic compact stars in $f(R,G)$ gravity:\\
\begin{itemize}
 \item The EOS parameters for compact stars are true as in the case of ordinary matter distribution in $f(R,G)$ gravity which shows that the compact stars are composed of ordinary matter. The matter density, and radial and tangential pressures get the maximum values at the center of the star and they are also decreasing functions. It confirms the fact that the matter components of compact stars are positive and remain finite in the interior of stars. Thus, all three compact stars in this present study are singularity free.
 \item  It is also noticed that anisotropic force is directed outward for $p_t> p_r$ which means $\triangle>0$, while on the other hand the anisotropic force is directed inward for $p_t< p_r$ which shows that $\triangle<0$ . The graphical description of $\triangle>0$ for three different compact stars is revealed in Fig. ($14$).
 \item The energy conditions and the TOV are analyzed for all three compact stars under consideration. The behavior of energy conditions and TOV are shown in Fig.($8$) and Fig.($10$)  for stars Her $X1$ and 4U respectively. Although these are verified for remaining two stars as well. Furthermore, the inequality $\mid \upsilon_{sr}^2-\upsilon_{st}^2\mid <1$ is satisfied as shown in Fig. ($13$) for the considered compact stars. So these models of  stars are stable.\\
 \end{itemize}
The important feature of the present study is the $3$-dimensional analysis and all the results show that the compact stars behave as usual for the $f(G)$ model parameter $0<n<200$. Thus it is conjectured that the stars behave in the same way for a polynomial form of $f(G)$, $n\geq200$. Also the behavior of the stars can be checked for negative values of $n$ as well and considering some other forms of $f(R)$ gravity model. It is worth mentioning that the findings of the current study are in conformity with the results in \cite{31} for $n=2$ and $f_1(R)=0$.\\\\

\end{document}